\documentclass{article}

\usepackage{graphicx}
\usepackage{feynmf}

\newcommand\mysection{\setcounter{equation}{0}\section}

\newcounter{hran}

\begin{document}

\begin{fmffile}{samplepics}
\setlength{\unitlength}{1mm}

\begin{titlepage}

\begin{flushright}
CERN--TH/2002-017\\
LPT-Orsay 02 - 24\\
LAPTH-907/02\\
\end{flushright}
\vspace{1.cm}

\begin{center}

{\large \bf CROSS SECTION OF ISOLATED PROMPT PHOTONS IN HADRON--HADRON 
COLLISIONS}\\[2cm]

{\large S. Catani$^{a}$\footnote{On leave of absence from INFN, Sezione di
Firenze, Florence, Italy.}, M. Fontannaz$^{b}$, J.~Ph.~Guillet$^{c}$ and  
E.~Pilon$^{c}$} \\[.5cm]

\normalsize
{$^{a}$ Theory Division, CERN, CH-1211 Geneva 23, Switzerland}\\[.2cm] 
{$^{b}$ Laboratoire de Physique Th\'eorique LPT,}\\
{Universit\'e de Paris XI, F-91405 Orsay Cedex, France}\\[.2cm]        
{$^{c}$ Laboratoire d'Annecy-Le-Vieux de Physique Th\'eorique LAPTH,}\\
{F-74941 Annecy-le-Vieux, France}\\
      
\end{center}

\vspace{2cm}

\begin{abstract} 
\noindent
We consider the production of isolated prompt photons in hadronic collisions. 
We present a general discussion in QCD perturbation theory of the 
isolation criterion used by hadron collider experiments. The isolation
criterion is implemented in a computer programme of the Monte Carlo type, 
which evaluates the production cross section at next-to-leading order accuracy
in perturbative QCD. The calculation includes both the direct and the
fragmentation components of the cross section, without any approximation
of the dependence on the radius $R$ of the 
isolation cone. 
We examine the scale dependence of the isolated cross section, 
the sensitivity of the cross section to the values of the 
isolation parameters,  and we
provide a quantitative comparison between the full $R$ dependence and its
small-$R$ approximation. \end{abstract}

\vspace{3cm}

\vspace*{\fill}
\begin{flushleft}
     CERN--TH/2002-017 \\  
     February 2002
\end{flushleft}

\end{titlepage}

\pagestyle{plain} 

\mysection{Introduction}\label{intro} 

The production of prompt photons with large transverse momenta $p_{T}$ at 
hadronic colliders has been  the subject of a continuing effort, both
experimentally \cite{UA1,UA2,cdf,d0,cdfnew,HERA} and theoretically 
\cite{abfs,abf,owens,berger-qiu,gordon-vogelsang2,vogt-vogelsang},
during the last fifteen years. The expression `prompt photons' means that these
photons do not come from the decay of hadrons, such as $\pi^{0}$, $\eta$, etc., 
produced at large transverse momenta. Prompt-photon production is an interesting
observatory of the short-distance dynamics of quarks and gluons by using a hard
colourless probe.
It is complementary to the electroweak processes of Deep Inelastic Scattering 
and Drell--Yan pair production, and to pure QCD processes such as inclusive 
production of jets or heavy flavours.
One of the main motivations of its study is its sensitivity
to the gluon density inside the colliding hadrons \cite{abfow}. Indeed the
gluon distribution function is involved already at the lowest order (LO) 
in the strong coupling $\alpha_s$
through the `QCD Compton' subprocess $q g \to \gamma q$, which dominates at 
fixed-target
energies. Moreover the point-like coupling of the photon to quarks in principle
makes the process of prompt-photon production ideally free from the
uncertainties inherent in jet reconstruction (as in the case of jet
production) or in fragmentation of partons into hadrons (as in inclusive
hadron production). 

\vspace{0.3cm}

\noindent 
In fact, the latter feature is not as ideal as initially thought, because prompt
photons can be produced according to two possible mechanisms, one of them being
a fragmentation mechanism. Whereas the contribution from fragmentation
remains small (less than 10\%) at fixed-target energies, it becomes dominant in
inclusive prompt-photon production at colliders. This was already true at the
CERN $Sp\bar{p}S$, at least in the lower range of the $p_{T}$ spectrum for the
UA1 and UA2 experiments at the centre-of-mass (c.m.)
energy $\sqrt{S}$ = 630 GeV. It is true at the Tevatron ($\sqrt{S}\sim 2$~TeV), 
and will remain true at the LHC ($\sqrt{S}= 14$~TeV), since
this dominance increases with $\sqrt{S}$.

\vspace{0.3cm}

\noindent
To be precise,
the collider experiments at the Tevatron and at the forthcoming
LHC do not perform {\it inclusive} photon measurements. The background of 
secondary photons coming from the decays of $\pi^{0}$, $\eta$, etc., overwhelms the 
signal by several orders of magnitude. To reject this background, the
experimental selection of prompt photons requires {\em isolation} cuts. The 
isolation criterion used by collider experiments is schematically as follows. 
A photon is said to be isolated if, in a cone of radius $R$
in rapidity and 
azimuthal angle around the photon direction, the amount of deposited hadronic 
transverse energy $E_{T}^{had}$ is smaller than some value 
$E_{T \, max}$ 
chosen by the experiment:  
\begin{equation}\label{criterion}     
E_{T}^{had} \leq E_{T \, max} \;\;\;\; \mbox{inside} \;\;\;\;      
\left( y - y_{\gamma} \right)^{2} +    
\left(  \phi - \phi_{\gamma} \right)^{2}  \leq R^{2}  \;.    
\end{equation}     
In addition to the rejection of the background of secondary  photons, the 
isolation cuts also affect the prompt-photon cross section itself, 
in particular by reducing the effect of fragmentation.

\vspace{0.3cm}

\noindent
Hadronic production of isolated prompt photons is not relevant only to QCD
studies.
The LHC experiments, ATLAS and CMS, will search neutral Higgs 
bosons in the mass range 80--140~GeV through the decay channel $H \to \gamma \gamma$
\cite{atlascms}. 
The production of pairs of isolated prompt photons with a large 
invariant mass is the so-called irreducible background to this search.
Therefore, a quantitative understanding of
this background is relevant to reliably estimate the expected
significance $S/\sqrt{B}$ ($S$ and $B$ being the number of expected  Higgs
boson events and background events, respectively). This provides another
important motivation for studying
isolation in prompt-photon reactions. 

\vspace{0.3cm}

\noindent
Various
issues on isolation of photons based on the above criterion\footnote{Other 
isolation criteria have also been discussed. We mention, in
particular, the alternative criterion proposed in ref.~\cite{frixione}
and studied in ref.~\cite{Frixione:2000gr} and in sect.~6.3 of 
ref.~\cite{Catani:2000jh}. 
The related topic of isolated photons produced in $e^{+}e^{-}$ collisions
has also been abundantly discussed. A variant of (\ref{criterion})
suitable for $e^{+}e^{-}$ was studied in \cite{kt} and revisited in
\cite{berger-guo-qiu,Aurenche:1997ng,cfp}. A different criterion \cite{glover}
for isolated photons in jets
has also been applied in measurements of 
the LEP
experiments.}\label{f1} have already been discussed in the literature
\cite{abf,owens,berger-qiu,gordon-vogelsang2,vogt-vogelsang}. The aim of the
present article is to provide a detailed QCD study of the isolation criterion 
(\ref{criterion}), and to present a complete (without any small-$R$
approximation) calculation at 
next-to-leading order (NLO) accuracy in QCD perturbation theory.

\vspace{0.3cm}

\noindent
The first theoretical question raised by the implementation of isolation is an 
issue of principle. Since the production of isolated photons, and more
generally of any isolated particle, is no longer an inclusive process, one may
question the validity of the factorization theorem of collinear singularities
that is established in the inclusive case \cite{berger-guo-qiu}. 
In the case of $e^{+}e^{-}$ annihilation into an isolated prompt photon plus
hadrons, this issue was examined in ref.~\cite{cfp}. It was shown that
factorization still holds in terms of the same fragmentation functions as
appear in the inclusive case, whereas the dependence on the isolation parameters
is consistently taken into account by the short-distance partonic cross section.
The case of hadronic collisions is more involved, since factorization 
is complicated by the presence of initial-state collinear singularities in the 
perturbative calculation. 
In this paper we extend the treament of ref.~\cite{cfp} to hadronic collisions,
we consider the isolation criterion (\ref{criterion}) and some variants of it,
and we discuss and prove factorization to any order in QCD perturbation theory.

\vspace{0.3cm}

\noindent
The evaluation of NLO QCD corrections to the inclusive production of a single
isolated prompt photon in hadron--hadron collisions was considered in 
refs.~\cite{owens} and \cite{gordon-vogelsang2}. The authors of 
ref.~\cite{owens} computed the NLO corrections to direct photon production,
without including those to the fragmentation mechanism.
The NLO calculation of ref.~\cite{gordon-vogelsang2} includes both the direct
and fragmentation mechanisms, and it uses the small-cone approximation
$R \ll 1$.
More precisely, the approach of ref.~\cite{gordon-vogelsang2} is based on 
an analytical treatment of isolation in which the transverse
energies, rapidities and azimuthal angles in eq.~(\ref{criterion}) are replaced
by energies and relative angles defined in the hadronic c.m. frame.
This substitution of variables has no effect in the small-cone approximation.
To go beyond the small-cone approximation, 
in this paper we carry on our study in terms of the variables
$\eta, \phi$ and $E_{T}$ involved in the criterion (\ref{criterion}). 
Since these variables are invariant under longitudinal boosts
along the direction of the beam axis, the criterion in eq.~(\ref{criterion})
is current practice in experiments at high-energy hadron colliders.
We perform a complete NLO calculation: it includes both the direct and 
fragmentation contributions without any small-cone approximation.
Our approch is based on a combined analytical and numerical treatment,
and it can be used to implement different variants of the isolation criterion 
(\ref{criterion}).

\vspace{0.3cm}

\noindent
This paper
is organized as follows. In sect.~\ref{theory}, we 
recall the two production mechanisms (direct and fragmentation)
of a prompt photon. In sect.~\ref{ancalc_incl} we present a brief pedagogical
discussion on non-isolated prompt photons: using the collinear approximation,
we derive analytic expressions for the direct and fragmentation contributions 
to the NLO cross section. The general factorization issues that
are raised in the case of isolated
particle hadroproduction are discussed in sect.~\ref{factorisation}.
In sect.~\ref{ancalc_isol}, we show how the implementation of isolation affects
the analytic expressions of sect.~\ref{ancalc_incl}
in the limit of small values of the radius $R$ of the isolation cone.
To go beyond the small-$R$ approximation, we implement the NLO
calculation of the isolated-photon cross section 
in a computer programme of the
Monte Carlo type. The computer programme, which enables us to evaluate the
full $R$ dependence of the cross section, is described in sect.~\ref{numcalc},
where we also present some numerical results and comparisons with the 
small-$R$ approximation.
In sect.~\ref{conclusion} we summarize our results and we conclude
with some remarks on questions left open when the cone radius $R$ becomes small
or the isolation condition becomes very tight (when $E_{T \,  max}$ becomes small).

\mysection{Mechanisms of prompt-photon production}\label{theory}

As mentioned in sect.~\ref{intro}, schematically, a high-$p_{T}$ prompt photon 
can be produced by two possible mechanisms: either it takes part 
directly in the hard subprocess, or it results from the collinear fragmentation
of a parton that is itself produced with a large transverse momentum. From a
topological point of view,
when a `direct' photon is produced, it is most probable that it will be
separated from the hadronic environment, whereas a photon `from fragmentation'
is most probably accompanied by hadrons, except when the photon carries away
most of the momentum of the fragmenting parton. These fragmentation
configurations are rare and atypical, and they are
precisely those that are not suppressed by the isolation criterion. 

\vspace{0.2cm}

\noindent
The LO contribution to direct prompt-photon production is given by the Born-level
processes $q \bar{q} \rightarrow \gamma g$ and  $q$ (or $\bar{q}$) $g$
$\rightarrow \gamma q$ (or $\bar{q}$). The computation of the NLO contributions 
yields ${\cal O}(\alpha_s)$ corrections coming from the subprocesses 
$q \bar{q} \rightarrow \gamma g g$, $g q$ (or $\bar{q}$) 
$\rightarrow \gamma g q$ (or $\bar{q}$) and from the virtual corrections
to the Born-level processes. 

\vspace{0.3cm}

\noindent
The calculation of these higher order corrections also yields the LO 
contribution of the fragmentation type (sometimes called `bremsstrahlung 
contribution'), in which the photon comes from the collinear fragmentation of a
hard parton produced in a short-distance subprocess 
(see, for instance, Diagram~a). 
This contribution appears in the following way. A final-state quark--photon 
collinear singularity occurs in the calculation of the contribution from the 
subprocess $g q \rightarrow \gamma \gamma q$. 
At higher orders, final-state multiple collinear singularities appear in any
subprocess where a high-$p_{T}$ parton (quark or gluon) undergoes a cascade of
successive collinear splittings ending up with a quark--photon splitting. These
singularities are factorized to all orders in $\alpha_s$ 
and absorbed into quark and gluon fragmentation
functions of the photon, $D^{\gamma}_q(z,M_{F}^{2})$ and 
$D^{\gamma}_g(z,M_{F}^{2})$, defined in a certain factorization
scheme at a factorization scale $M_{F}$ 
chosen to be of the order of the hard scale of the process.
When the fragmentation scale $M_{F}$
is large with respect to $\sim 1 $~GeV, these functions behave
roughly as $\alpha/\alpha_s(M_{F}^{2})$, so that these contributions
are 
of the same order in $\alpha_s$ as the Born-level terms in the direct mechanism. 
Moreover, because of the high value of the gluon parton densities
at small momentum fraction  $x$, the $g q$ (or $\bar{q}$) initiated
contribution producing one photon from fragmentation even dominates the
inclusive production rate in the lower range of the photon $p_{T}$-spectrum at the
Tevatron, and this will be even more true at the LHC. The calculation of the NLO
corrections to the fragmentation contribution 
(see, for instance, Diagrams~b and c) is thus required for a reliable and
consistent treatment of prompt-photon production at this accuracy.
\begin{figure}
\begin{center}
\[
\parbox[c][60mm][c]{40mm}{\begin{fmfgraph*}(40,40)
  \fmfleft{H1,H2}
  \fmfright{Hf1,qq,ph1,ph2,Hf2}
  \fmf{dbl_plain_arrow,label.side=right,tension=2,lab=$\bar{p}$}{H1,VP1}
  \fmf{dbl_plain_arrow,label.side=right,tension=2,lab=$p$}{H2,VP2}
  \fmf{plain}{VP1,Hf1}
  \fmf{plain}{VP2,Hf2}
  \fmfblob{.16w}{VP1,VP2}
  \fmf{gluon,tension=1.5}{VP2,v1}
  \fmf{fermion}{VP1,v1,v2,vi,qq}
  \fmf{gluon}{v2,ph2}
  \fmfv{label=Diagram a,label.angle=-90.,label.dist=0.3w}{VP1}
  \fmffreeze
  \fmf{photon}{vi,ph1}
  \fmfi{plain}{vpath (__VP1,__Hf1) shifted (thick*(0,2))}
  \fmfi{plain}{vpath (__VP1,__Hf1) shifted (thick*(0,-2))}
  \fmfi{plain}{vpath (__VP2,__Hf2) shifted (thick*(0,2))}
  \fmfi{plain}{vpath (__VP2,__Hf2) shifted (thick*(0,-2))}
\end{fmfgraph*}}
\qquad + \cdots
\]
\end{center}
\end{figure}
%
%

\begin{figure}
\begin{center}
\[
\parbox[c][60mm][c]{40mm}{\begin{fmfgraph*}(40,40)
  \fmfleft{H1,H2}
  \fmfright{Hf1,ph1,ph2,Hf2}
  \fmf{dbl_plain_arrow,label.side=right,tension=2,lab=$\bar{p}$}{H1,VP1}
  \fmf{dbl_plain_arrow,label.side=right,tension=2,lab=$p$}{H2,VP2}
  \fmf{plain}{VP1,Hf1}
  \fmf{plain}{VP2,Hf2}
  \fmfblob{.16w}{VP1,VP2}
  \fmf{gluon,tension=1.5}{VP2,v1}
  \fmf{fermion}{VP1,vi1}
  \fmf{fermion}{vi1,v1}
  \fmf{fermion}{v1,v2}
  \fmf{fermion,tension=.1}{v2,vi2}
  \fmf{fermion}{vi2,vi}
  \fmf{photon}{vi,ph1}
  \fmf{gluon}{v2,ph2}
  \fmfblob{.10w}{vi}
  \fmfv{lab=$D_{\gamma/q}$,lab.dist=.1w,lab.ang=-90}{vi}
  \fmf{gluon,tension=.5}{vi1,vi2}
  \fmfv{label=Diagram b,label.angle=-90.,label.dist=0.3w}{VP1}
  \fmffreeze
  \fmf{gluon}{vi1,vi2}
  \fmfi{plain}{vpath (__VP1,__Hf1) shifted (thick*(0,2))}
  \fmfi{plain}{vpath (__VP1,__Hf1) shifted (thick*(0,-2))}
  \fmfi{plain}{vpath (__VP2,__Hf2) shifted (thick*(0,2))}
  \fmfi{plain}{vpath (__VP2,__Hf2) shifted (thick*(0,-2))}
\end{fmfgraph*}}
\qquad + \cdots + \qquad
\parbox[c][60mm][c]{40mm}{\begin{fmfgraph*}(40,40)
  \fmfleft{H1,H2}
  \fmfright{Hf1,ph1,ph2,qq,Hf2}
  \fmf{dbl_plain_arrow,label.side=right,tension=2,lab=$\bar{p}$}{H1,VP1}
  \fmf{dbl_plain_arrow,label.side=right,tension=2,lab=$p$}{H2,VP2}
  \fmf{plain}{VP1,Hf1}
  \fmf{plain}{VP2,Hf2}
  \fmfblob{.16w}{VP1,VP2}
  \fmf{gluon}{VP2,v1}
  \fmf{gluon}{VP1,v2}
  \fmf{fermion}{qq,vi1,v1,v2,vi}
  \fmf{photon}{vi,ph1}
  \fmfblob{.10w}{vi}
  \fmfv{label=Diagram c,label.angle=-90.,label.dist=0.3w}{VP1}
  \fmffreeze
  \fmf{gluon}{vi1,ph2}
  \fmfv{lab=$D_{\gamma/q}$,lab.dist=.08w,lab.ang=-90}{vi}
  \fmfi{plain}{vpath (__VP1,__Hf1) shifted (thick*(0,2))}
  \fmfi{plain}{vpath (__VP1,__Hf1) shifted (thick*(0,-2))}
  \fmfi{plain}{vpath (__VP2,__Hf2) shifted (thick*(0,2))}
  \fmfi{plain}{vpath (__VP2,__Hf2) shifted (thick*(0,-2))}
\end{fmfgraph*}}
\qquad + \cdots
\]

\end{center}
\end{figure}

\vspace{0.3cm}

\noindent
In this paper, we call `direct' the contribution given by the Born term plus the
part of
the higher-order corrections from which final-state collinear singularities
have been subtracted according to the ${\overline{MS}}$ factorization scheme.
The remaining part is called `fragmentation' contribution, and involves the fragmentation function 
of a parton into a photon as defined in the ${\overline{MS}}$ factorization scheme.
In our numerical calculations, we use the fragmentation functions of
ref.~\cite{bfg}.

\vspace{0.3cm}

\noindent 
Note, however, that the above distinction between the two mechanisms has no
direct physical meaning beyond LO.
From a theoretical point of view, the distinction is defined by an 
arbitrary choice. It follows from the necessity of factorizing the final-state 
collinear singularities and absorbing them into the fragmentation functions.
This factorization requires the introduction of an arbitrary 
fragmentation scale $M_{F}$, which is an unphysical parameter. More generally, 
it relies on an arbitrary choice of the factorization scheme, which 
defines the finite part of the higher-order corrections that is
absorbed in the fragmentation functions together with the singularities; 
the remaining finite part is then included in the higher-order
contributions to the partonic cross sections.
The dependence on this
arbitrariness, and in particular on $M_{F}$, cancels only in the sum of the
direct and fragmentation contributions, so only this sum is a physical observable. 
In particular, any experimental identification of direct and fragmentation
contributions based on topological grounds does not match the `theoretical
identification' used throughout this paper.

\mysection{Inclusive cross section}\label{ancalc_incl}

We start the discussion by considering the inclusive cross section for the
production of a non-isolated photon with momentum $p_{\gamma}$. The transverse
momentum and the rapidity of the photon are denoted by 
$p_{ T \, \gamma}$ and $y_{\gamma}$ (since the photon is massless, 
$y_{\gamma}=\eta_{\gamma}$ where $\eta_{\gamma}$ is the pseudorapidity),
respectively. The hadronic cross section $d\sigma/dp_{ T \, \gamma}
dy_{\gamma}$, which we denote by $\sigma(p_{\gamma})$,
is given by the sum of 
the `fragmentation' and `direct' contributions. 
It can be written, in shorthand, as
\begin{equation}
\label{2.1} 
\sigma(p_{\gamma}) = \sum_{a} \int_0^1 \frac{dz}{z} \;
\widehat{\sigma}^{a}(p_{\gamma}/z;\mu,M,M_{F}) D_{a}^{\gamma}(z;M_F)
+ \widehat{\sigma}^{\gamma}(p_{\gamma};\mu,M,M_{F}) \;,
\end{equation}
where $\widehat{\sigma}^{a}$ and $\widehat{\sigma}^{\gamma}$ are the
corresponding `partonic' cross sections.
The contribution $\widehat{\sigma}^{a}$ describes the production of a parton
$a$ ($a=q,{\bar q},g$) in the hard collision, 
and $D_{a}^{\gamma}$ is the fragmentation function of the parton $a$
into a photon. The direct contribution 
$\widehat{\sigma}^{\gamma}$ does not contain any fragmentation function;
it corresponds to the
point-like coupling of the large-$p_{T}$ photon to a quark produced in the hard
subprocess. Note that $\widehat{\sigma}^{a}$ and $\widehat{\sigma}^{\gamma}$
are not true partonic cross sections, since they include the convolution
with the parton distributions of the colliding hadrons.

\vspace{0.3cm}

\noindent
The cross sections 
$\widehat{\sigma}^{\gamma}$ and $\widehat{\sigma}^{a}$ are known
up to NLO in $\alpha_{s}$:
\begin{eqnarray}
\widehat{\sigma}^{\gamma}(p;\mu ,M,M_{F}) & = & 
\left ( \frac{\alpha_s(\mu )}{\pi} \right ) \;
\sigma_{Born}^{\gamma}(p;M) + 
\left( \frac{\alpha_s(\mu)}{\pi} \right) ^{2}
\sigma_{HO}^{\gamma}(p;\mu,M, M_{F}) \label{2.2a} \;,\\
 \widehat{\sigma}^{a}(p;\mu,M,M_{F}) & = & 
\left( \frac{\alpha_s(\mu)}{\pi} \right) ^{2}
\sigma^{a}_{Born}(p;M) + 
\left( \frac{\alpha_s(\mu)}{\pi} \right) ^{3} 
\sigma^{a}_{HO}(p;\mu,M,M_F) \;. \label{2.2b}
\end{eqnarray} 
The expressions of $\sigma_{HO}$ (and $\sigma_{Born}$) for the direct and
fragmentation contributions
can be found in refs.~\cite{abfs,gordon-vogelsang2} and ref.~\cite{acgg},
respectively.
They depend on the renormalization scale $\mu$, on the factorization
scale $M$ of the initial-state parton distributions and on the
factorization scale $M_{F}$ of the photon fragmentation function.
These results have
been the starting point of the prompt-photon phenomenology of the last fifteen
years \cite{abfs}. More recently, expressions involving resummation of
logarithmic terms that are large at the phase-space boundary also became
available \cite{los,cmn,cmnov,Kidonakis:1999hq,sv,Laenen:2001ij}.

\vspace{0.3cm}

\noindent
To better understand the differences between the inclusive cross
section and the isolated cross section, we need a more detailed expression of
$\sigma(p_\gamma)$ in eq.~(\ref{2.1}). Let us first consider the fragmentation 
contribution in the Born approximation:
\begin{eqnarray}
d\sigma^{brems}_{Born}[{\scriptstyle A+B \to \gamma + jet + X}] 
& = &
\frac{1}{8 \pi S^{2}} \sum_{i,j\atop{a,b}} \int \ 
    \frac{G_{i/A}(x_{1}, M)}{x_{1}} \ \frac{G_{j/B}(x_{2}, M)}{x_{2}} \
    |\overline{\cal M}_{B}[{\scriptstyle i+j \to a+b}]|^{2} \nonumber \\
& & \;\;\;\;\;\;\;\;\;\;\;\;\;\;\; \times \;\;
    D_{a}^{\gamma}(z,M_F) \ \Theta \left( z - z_{min} \right) \ dz \
   d\eta_{b} \ d\eta_{a} \ p_{ T \, a} \ dp_{ T \, a} . \label{2.3}
\end{eqnarray} 

\noindent 
Here we are considering the photon--jet double inclusive cross section. The
photon has transverse momentum $p_{ T \, \gamma } = z \, p_{ T \, a }$ and 
rapidity $\eta_{\gamma} = \eta_{a}$. The jet recoiling against the photon, 
here the parton $b$, has rapidity $\eta_{jet} = \eta_{b}$. 
$|\overline{\cal M}_{B}|^{2}$ is the matrix element squared of the 
Born subprocess, averaged (summed) over initial-state (final-state) spins 
and colours. $G_{i/A}(x_{1},M)$ and $G_{j/B}(x_{2},M)$ are the parton 
distribution functions of the incoming hadrons, with $x_{1,2}$ given 
in the 
hadronic c.m. frame by
\begin{equation}\label{2.4}
x_{1\atop2} = \frac{p_{ T \, a }}{\sqrt{S}} \ 
\left( e^{\pm \eta_{a}} + e^{\pm \eta_{b}} \right) \;,
\end{equation}

\noindent 
where transverse momenta, rapidities and the hadronic energy
$\sqrt{S}$ are defined in the c.m. frame. We rewrite (\ref{2.3}) in a 
more compact form:
\begin{eqnarray}
d\sigma^{brems}_{Born}[{\scriptstyle A + B \to \gamma + jet + X}] & = & 
2 \pi \left( \frac{\alpha_s(\mu)}{\pi} \right) ^{2}  
    \sum_{a,b} \int_{z_{min}}^{1} D_{a}^{\gamma}(z,M_{F}) \ dz \nonumber \\
& & \times \;\; \int E_{a} 
    \frac{d\widehat{\sigma}_{Born}[{\scriptstyle A + B \to a + b + X}]}
         {d\vec{p}_{a} \ d\eta_b} \ d\eta_{b} \ d\eta_{a} \ p_{ T \, a } \ 
	 dp_{ T \, a } \;,\label{2.5}
\end{eqnarray}

\noindent 
where $z_{min}$ is given by
\begin{equation}\label{2.6}
z_{min} =  max \left\{ {p_{ T \, \gamma } \over \sqrt{S}} \left ( e^{\pm
\eta_{\gamma}} + e^{\pm \eta_b} \right ) \right\}  \;.
\end{equation}

\vspace{0.2cm}

\noindent 
We can write a similar expression for the higher-order (HO) corrections to the
fragmentation contribution: 
\begin{eqnarray}
\lefteqn{d\sigma^{brems}_{HO}[{\scriptstyle A+B \to \gamma + jet +x}] }
\nonumber \\ 
& = & 2 \pi \left( \frac{\alpha_{s}(\mu)}{\pi} \right) ^{3} 
      \sum_{a} \int E_{a} 
      \frac{d\widehat{\sigma}_{HO}[{\scriptstyle A + B \to a + jet + X}]}
           {d\vec{p}_{a} \ d\eta_{jet}} \ D_a^{\gamma}(z,M_F) \ 
      dz \ d\eta_{jet} \ d\eta_a \ p_{ T \, a } \ dp_{ T \, a }  \\ 
& = & 2 \pi \left ( \frac{\alpha_{s}(\mu)}{\pi} \right) ^{3}
      \sum_{a,b,i} \int_{z_{min}}^{1} dz \ D_i^{\gamma}(z, M_F)  
      \int_{\frac{z_{min}}{z}}^{1} dx \ P_{ia}(x) \
      \ln \left( {p_{T \gamma} \over M_{F}} \right) \nonumber \\
&   & \;\;\;\;\;\;\;\;\;\;\;\;\;\;\;\;\;\;\;\;\;\;\;\;\;\;\;\; 
      \times \;\;\; E_{a} 
      \frac{d\widehat{\sigma}_{Born}[{\scriptstyle A + B \to a + b + X}]} 
          {d\vec{p}_{a} \ d\eta_{b}} \ 
      d\eta_{b} \ d\eta_{a} \ p_{ T \, a } \ dp_{ T \, a }  \label{2.7} \\
& + & 2 \pi \left( \frac{\alpha_{s}(\mu)}{\pi} \right) ^{3} 
      \sum_a \int dz \ D_{a}^{\gamma}(z,M_{F}) \
      K^{brems}_{HO}[{\scriptstyle A + B \to a + X}]( p_{ T \, \gamma },\mu,M)
      \;. \nonumber
\end{eqnarray}  

\noindent
In eq.~(\ref{2.7}), the $M_{F}$ dependence of the HO corrections has been made
explicit. The remainder of the HO corrections is given by $K^{brems}_{HO}$ and 
no longer 
depends on $M_{F}$. The kernels $P_{ia}(x)$ are the Altarelli--Parisi 
splitting functions.

\vspace{0.3cm}
\noindent
The direct contribution has the following form:

\begin{eqnarray}
\lefteqn{d\sigma^{dir}[{\scriptstyle A + B \to \gamma + jet + x}] } \nonumber \\
&  = & 2 \pi \left( \frac{\alpha_s(\mu)}{\pi} \right) \;\;
       \left(\frac{\alpha}{\pi} \right) \sum_{b} E_{\gamma} 
       \frac{d\widehat{\sigma}_{Born}[{\scriptstyle A + B \to \gamma +b}]}
            {d \vec{p}_{\gamma} \ d\eta_{b}} d\eta_{b} \ d\eta_{\gamma} 
       \ p_{ T \, \gamma } \ dp_{ T \, \gamma } \nonumber \\
& + &  \ 2 \pi  \left( \frac{\alpha_{s} (\mu)}{\pi} \right) ^{2}
       \left( \frac{\alpha}{\pi} \right)
       \sum_{a,b\atop{a= q, \bar{q}}} \int_{z_{min}}^{1} 
       dz \ e_{a}^{2} \ P_{\gamma q}(z) \ 
       \ln \left( \frac{p_{ T \, \gamma }}{M_F} \right) \nonumber \\      
&   &  \;\;\;\;\;\;\;\;\;\;\;\;\;\;\;\;\;\;\;\;\;\;\;\;\;\;\;\;\;\;\;\;\;\;\;
       \times \;\;\; E_{a} 
       \frac{d \widehat{\sigma}_{Born}[{\scriptstyle A + B \to a + b}]}
            {d\vec{p}_{a} \ d\eta_{b}} d\eta_{b} \ d\eta_{a} \ p_{T \, a} \ 
      dp_{T \, a} \nonumber \\ 
& + & 2 \pi \left( \frac{\alpha_{s}(\mu)}{\pi} \right) ^{2} 
      \left( \frac{\alpha}{\pi} \right)
      \ K^{dir}_{HO}[{\scriptstyle A + B \to \gamma + jet + X}]
      (p_{ T \, \gamma },\mu,M) \;.\label{2.8}
\end{eqnarray}

\noindent Here we have included QCD corrections up to NLO and 
also made the $M_{F}$ dependence of the HO corrections explicit. 
The splitting function $P_{\gamma q}(z)$ of a quark $q$ into a
photon is
\begin{equation}
P_{\gamma q}(z) = \frac{1 + (1 - z)^2}{z} \;.
\label{2.9}
\end{equation}

\noindent
The LO evolution equations satisfied by the fragmentation functions are
\begin{equation}\label{2.10e}
M_{F}^{2} \frac{\partial D_a^{\gamma}(z,M_{F})}{\partial M_{F}^{2}} = 
\frac{\alpha}{2 \pi} \ e_{a}^{2} \ P_{\gamma q}(z) +
\frac{\alpha_{s}}{2 \pi} \left[ \sum_q \left( P_{qa} + P_{\bar{q}a} \right) 
\otimes D_{q}^{\gamma}(M_{F}) + 
P_{ga} \otimes D_{g}^{\gamma}(M_{F}) \right] ,
\end{equation}
where the symbol $\otimes$ denotes convolutions with respect to the momentum
fraction $z$, and $e_{a}$ is the electric charge of the parton $a$.
Using eq.~(\ref{2.10e}), it is straightforward to verify
the $M_{F}$-independence of the prompt-photon cross section up to 
NLO accuracy. 

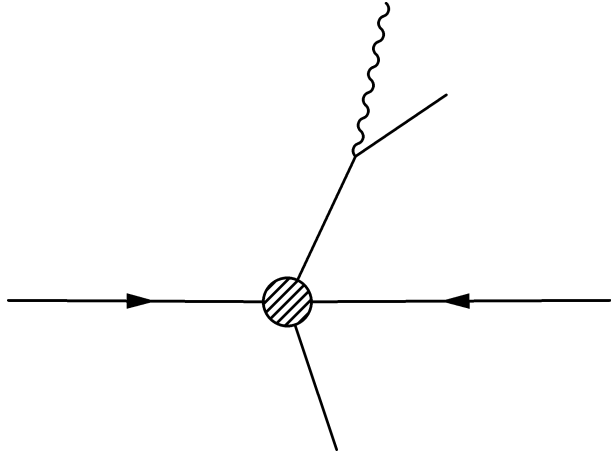
\begin{figure}[p]
\begin{center}
\parbox[c][80mm][c]{80mm}{\begin{fmfgraph*}(80,80)
  \fmfleftn{i}{1} \fmfrightn{o}{1}
  \fmftopn{t}{9} \fmfbottomn{b}{9}
  \fmf{fermion,label=1,label.side=left,tension=2.}{i1,v1}
  \fmf{fermion,label=2,tension=1.}{o1,v1}
  \fmf{photon,tension=1.5}{t6,v2}
  \fmf{plain,label=a,tension=2.}{v1,v2}
  \fmf{plain,label=b,tension=2.}{v1,fv1}
  \fmf{phantom,tension=2.}{fv1,b6}
  \fmfv{decor.shape=circle,decor.filled=shaded,decor.size=.08w}{v1}
  \fmf{plain,tension=1.}{v2,fv2}
  \fmfv{label=q,label.angle=-50.,label.dist=0.03w}{fv2}
  \fmf{phantom,tension=1.}{fv2,t8}
\end{fmfgraph*}}
\end{center}
\caption{\label{Fig:kinematics}{\em Kinematics of HO corrections to the 
direct contribution.}}
\end{figure}

\vspace{0.3cm}

\noindent
A short description of the derivation of eq.~(\ref{2.8}) is instructive and
helps to understand how the isolated cross section (see sect.~\ref{ancalc_isol})
can be obtained from the inclusive cross
section. Let us consider the
subprocess $1 + 2 \to \gamma + q + b$, in which the photon is emitted by the
final-state quark $a$ (fig.~\ref{Fig:kinematics}). Its contribution to
the cross section is 
\begin{eqnarray}\label{2.10} 
d\sigma 
& = & \int dx_1 \ G_{1/A}(x_1) \ dx_2 \ G_{2/B}(x_2) \nonumber \\
&   & \times \;\;
      \frac{1}{2 \widehat{s}} |\overline{\cal M}|^{2} 
      \widetilde{dp}_{\gamma} \ \widetilde{dp}_{q} \ \widetilde{dp}_{b} \
      (2 \pi)^{n} \ 
      \delta^{(n)} \left( p_{1} + p_{2} - p_{b} - p_{\gamma} - p_{q} \right) 
      \;\;, 
\end{eqnarray}

\noindent 
where we use the definition 
$\widetilde{dp} \equiv d^{n}p \ \delta^{(+)}(p^{2})/(2 \pi)^{n-1}$, and 
$n = 4 - 2 \varepsilon$ is the number of space-time dimensions. 
We are mainly interested in the singular contribution (non-collinear 
contributions are discussed in sect.~\ref{numcalc}) to 
eq.~(\ref{2.10}) coming from the configuration in which the final-state quark and the 
photon are collinear. Therefore the matrix element of the subprocess 
$1 + 2 \to \gamma + q + b$ can be approximated as 
\begin{equation}\label{2.11}
|\overline{\cal M}|^2 \simeq \frac{e_q^{2} \mu^{2 \varepsilon}}{p_{\gamma} \cdot p_{q}} 
\left( P_{\gamma q}(z) - \varepsilon z \right) 
|\overline{\cal M}_B|^{2} \;\;,
\end{equation}
where ${\cal M}_B$ is the Born-level amplitude of the subprocess 
$ 1 + 2 \to a + b$ in fig.~\ref{Fig:kinematics}, and we  
defined the longitudinal variable $z$ as follows:
\begin{equation}
\label{2.13}
z = {p_{ T \, \gamma } \over p_{ T \, \gamma } + p_{T \, q}} \;\;.
\end{equation} 

\begin{center}

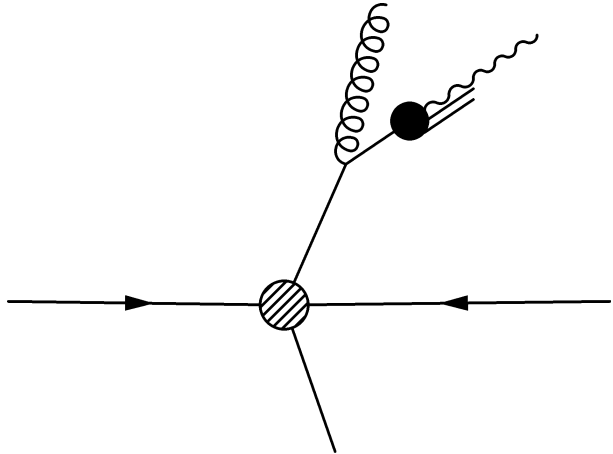
\begin{figure}[p]
\begin{center}
\parbox[c][80mm][c]{80mm}{\begin{fmfgraph*}(80,80)
  \fmfleftn{i}{1} \fmfrightn{o}{1}
  \fmftopn{t}{9} \fmfbottomn{b}{9}
  \fmf{fermion,label=1,label.side=left,tension=2.}{i1,v1}
  \fmf{fermion,label=2,tension=1.}{o1,v1}
  \fmf{gluon,tension=1.5}{t6,v2}
  \fmfv{label=5,label.angle=-50.,label.dist=0.03w}{t6}
  \fmf{plain,label=c,tension=2.}{v1,v2}
  \fmf{plain,label=b,tension=2.}{v1,fv1}
  \fmf{phantom,tension=2.}{fv1,b6}
  \fmfv{decor.shape=circle,decor.filled=full,decor.size=.06w}{fv2}
  \fmfv{decor.shape=circle,decor.filled=shaded,decor.size=.08w}{v1}
  \fmf{plain,label=a,label.side=right,tension=1.}{v2,fv2}
  \fmf{photon,tension=1.}{fv2,fv3,t8}
  \fmffreeze
  \fmfv{label=f,label.angle=-50.,label.dist=0.03w}{fv3}
  \fmfi{plain}{vpath (__fv2,__fv3) shifted (thick*(0,-2))}
  \fmfi{plain}{vpath (__fv2,__fv3) shifted (thick*(0,-4))}
\end{fmfgraph*}}
\end{center}
\caption{\label{Fig:2}{\em Kinematics of HO corrections to the fragmentation 
contribution.}}
\end{figure}
\end{center}

\vspace{0.3cm}

\noindent
Inserting eq.~(\ref{2.11}) in eq.~(\ref{2.10}), we obtain 
\begin{eqnarray}
\label{2.12}
& & d\sigma  = 
    \frac{V(n-2) e_q^{2} \mu^{2 \varepsilon}}{2^{3} (2 \pi)^{2n-3} S^{2}} 
    \int \frac{dx_{1}}{x_{1}} \ \frac{dx_{2}}{x_{2}} \ G(x_{1},M) \ G(x_{2},M) \ 
    \left( P_{\gamma q}(z) - \varepsilon z \right) 
    |\overline{\cal M}_B|^{2}
    \ p_{ T \, \gamma }^{n-3} \ dp_{T \, \gamma} \ 
    d\eta_{\gamma} \ d\eta_{b} \nonumber \\
& & \;\;\;\;\;\;\;\;\;\;\;\;\;\;\;\;\;\;\;\;\;\;\;\;\;\;\;\;\;
    \;\;\;\;\;\;\;\times \;\;\;
\delta \left( x_{1} - \frac{\sum_{i} p_{i}^{(\scriptstyle +)}}{\sqrt{S}} \right)
\delta \left( x_{2} - \frac{\sum_{i} p_{i}^{(\scriptstyle -)}}{\sqrt{S}} \right) 
\frac{d^{n-2}p_{q} \ d \eta_{q}}{p_q \cdot p_{\gamma}} \;,
\end{eqnarray}

\noindent where $\eta_i$ are the pseudorapidities,  
$p_{T \, b} = |\vec{p}_{T \, q} + \vec{p}_{T \gamma}|$, and we have defined 
\[
\sum\limits_i p_i^{(\pm)} = p_{ T \, \gamma } e^{\pm \eta_{\gamma}} + 
p_{T \, q} e^{\pm \eta_q} + p_{T \, b} e^{\pm \eta_b} \;.
\]

\noindent The factor 
$V(n-2) = 2 \pi^{\frac{n-2}{2}}/\Gamma \left( \frac{n-2}{2} \right)$
comes from the integration over the azimuthal angle of the photon.
The phase space in eq.~(\ref{2.12}) can be written in terms of the
variable $z$ in eq.~(\ref{2.13}) and the difference $\phi_{q}$ between the 
azimuthal angles of the quark and the photon:
\begin{eqnarray}\label{2.14}
\lefteqn{\mu^{2 \varepsilon} \int \frac{d^{n-2}p_{q} \ d\eta_{q}}{p_{q} \cdot
p_{\gamma}}} \nonumber \\ 
  & = &  V(n-3) \mu^{2 \varepsilon}\int 
  \frac{(\sin \phi_{q})^{-2 \varepsilon} \ d\phi_{q} \ d\eta_{q}}
       {\cosh (\eta_{\gamma} - \eta_{q}) - \cos \phi_{q}}
     \ \frac{p_{T \, q}^{n-3} \ dp_{T \, q}}{p_{T \, q} \ p_{ T \, \gamma }} 
     \nonumber \\
& = & V(n-3) \left( \frac{\mu^{2}}{p_{ T \, \gamma }^{2}} \right) ^{\varepsilon}
      \int \frac{(\sin \phi_{q})^{-2 \varepsilon} d\phi_{q} d\eta_{q}}
                          {\cosh(\eta_{\gamma} - \eta_{q}) - \cos \phi_{q}} 
     \left( \frac{1-z}{z} \right) ^{n-4} \ \frac{dz}{z^{2}} \;.
\end{eqnarray}

\vspace{0.3cm}
\noindent
For the purpose of the discussion on isolated photons in
sect.~\ref{ancalc_isol},
we restrict the angular integration in eq.~(\ref{2.14})
to a cone of radius $R$ in the $\eta$-$\phi$ space around the photon. This
restriction does not change the
singular collinear term (proportional to the pole $1/\varepsilon$) we are
interested in. On the other hand, this produces a dependence on $\ln
R^{2}$ that we are going to use in sect.~\ref{ancalc_isol}. In the small-cone
approximation defined by the constraint  
$\Theta( R^{2} - \phi^{2}_{q} - (\eta_{\gamma} - \eta_{q})^{2})$ 
with $R^{2} \ll 1$, we can obtain a simple expression for the cross section. 
Using collinear expressions for the kinematic variables and defining the quark 
momentum $p_{a} = p_{\gamma} + p_{q}$, we have
\begin{eqnarray}\label{2.15}
p_{ T \, a }             & \simeq & \frac{p_{ T \, \gamma }}{z} \;,\nonumber \\
p_{ T \, b }             & \simeq & p_{T \, a}                 \;,\nonumber \\
\eta_q                 &    =   & \eta_{\gamma}             \;,\nonumber \\
\sum_{i} p_{i}^{(\pm)} & \simeq & \frac{p_{ T \, \gamma }}{z} 
                                  \left( e^{\pm \eta_{\gamma}} +
                                         e^{\pm \eta_{b}} \right) \;.
\end{eqnarray}

\noindent 
The integrations over $\eta_{q}$ and $\phi_{q}$ in (\ref{2.14}) are easily 
performed, and we finally obtain (neglecting terms of  
${\cal O}\left( R^{2}) \right)$
\begin{eqnarray}\label{2.17}
d\sigma^{cone} 
& = & \frac{V(n-2) p_{ T \, \gamma }^{n-3}}{2^{2} (2 \pi)^{n-2} S^{2}} 
      \int \frac{G_{1/A}(\widehat{x}_{1},M)}{\widehat{x}_{1}} \ 
           \frac{G_{2/B}(\widehat{x}_{2},M)}{\widehat{x}_{2}} \	  
      |\overline{\cal M}_{B}|^{2} p_{T \, a}^{n-3} \ dp_{ T \, a } 
      \ d\eta_{a} \ d\eta_{b} \nonumber\\
&   & \times \; \frac{\alpha}{2 \pi} e_{a}^{2} 
      \left( P_{\gamma q}(z) - \varepsilon z \right)       
      \left( - \frac{1}{\varepsilon} \right)
      \frac{\Gamma (1 - \varepsilon)}{\Gamma (1 - 2 \varepsilon)}      
      \left (\frac{4 \pi \mu^{2}}
                  {R^{2} p_{ T \, \gamma }^{2}(1-z)^{2}} \right) ^{\varepsilon} 
      dz \;\;,
\end{eqnarray} 

\noindent where the variables $\widehat{x}_{i}$ are given by 
\[
\widehat{x}_{{1 \atop 2}} = 
\frac{p_{ T \, \gamma }}{z \sqrt{S}} 
\left( e^{\pm \eta_{\gamma}} + e^{\pm \eta_{b}} \right) \;.
\]

\noindent
Equation (\ref{2.17}) can be rewritten as 
   
\begin{eqnarray}
\label{2.18}
d\sigma^{cone} & = &
2 \pi \left( \frac{\alpha_{s}(\mu)}{\pi} \right) ^{2} 
    \int_{z_{min}}^{1} E_{a} 
    \frac{d \widehat{\sigma}_{Born}({\scriptstyle A + B \to a + b})}
         {d\vec{p}_{a} \ d\eta_{b}}  
	 \ d\eta_{b} \ d\eta_{a} \  p_{T \, a}^{n-3} \ dp_{T \, a} \nonumber \\ 
& & \times \; \left\{ D_{a}^{\gamma}(z,M_{F}) + 
            \frac{\alpha}{2 \pi} e_{a}^{2} 
            \left[ 
                  2\ln \left( \frac{R \ p_{ T \, \gamma }\ (1-z)}
                                    {M_{F}} \right) 
                    P_{\gamma q}(z) + z 
            \right] 
	    \right\} dz  \;,
\end{eqnarray} 

\noindent where we have used the lowest-order definition
of the quark into a photon fragmentation function in the 
$\overline{MS}$ factorization scheme:

\begin{equation}
\label{2.19}
D_{q}^{\gamma}(z,M_{F},\varepsilon) \equiv - \frac{1}{\varepsilon} 
\frac{\Gamma(1 - \varepsilon)}{\Gamma(1 - 2 \varepsilon)} 
\left( \frac{4 \pi \mu^{2}}{M^{2}_{F}} \right) ^{\varepsilon} 
\frac{\alpha}{ 2 \pi} e_{q}^{2} P_{\gamma q}(z) \;.
\end{equation}

\vspace{0.3cm}

\noindent 
We clearly recognize two parts in eq.~(\ref{2.18}). The part proportional
to $D_{q}^{\gamma}(z,M_{F})$, which comes from the 
collinear pole ${1/\varepsilon}$ in eq.~(\ref{2.17}),
is nothing but the Born-level fragmentation component
in (\ref{2.3}). The remaining part contributes to the direct component
in eq.~(\ref{2.8}). Since it has been calculated in the collinear approximation, 
it only contains terms that are either proportional to $\ln R$ or constant
in the limit $R\to 0$. Among the constant terms there is the term 
proportional to $\ln (p_{T \, \gamma}/M_{F})$, which was explicitly written in 
(\ref{2.8}).

\vspace{0.3cm}

\noindent
The contribution to the direct component of 
the inclusive cross section coming from the region
outside the cone cannot be evaluated so straightforwardly, yet it must contain
a term proportional to $\ln (1/R)$ of the form
\begin{eqnarray}\label{outside} 
{d\sigma^{outside}}
& = &
2 \pi \left( \frac{\alpha_{s}(\mu)}{\pi} \right) ^{2} 
    \int_{z_{min}}^{1} E_{a} 
    \frac{d \widehat{\sigma}_{Born}({\scriptstyle A + B \to a + b})}
         {d\vec{p}_{a} \ d\eta_{b}} \nonumber \\ 
&   &
	\times \; \ d\eta_{b} \ d\eta_{a} \  p_{T \, a}^{n-3} \ dp_{T \, a} 
            \frac{\alpha}{2 \pi} e_{a}^{2} 
            \left[ 2\ln \left( \frac{1}{R} \right) P_{\gamma q}(z) 
            \right] 
	    dz \;.
\end{eqnarray} 

\noindent
Indeed, the $R$ dependence must cancel in the sum of the contributions from 
inside and from outside the cone. 

\vspace{0.3cm}

\noindent
The NLO calculation described so far is extended to the case of isolated photons in
sect.~\ref{ancalc_isol}. Before doing that, we discuss the validity of 
factorization of isolated photons at any order in QCD perturbation theory.

\def\bom#1{{\mbox{\boldmath $#1$}}}
\def\irlim{\raisebox{-1.2ex}{\rlap{\tiny $\;\;\;p_i \to 0$}}
\raisebox{0ex}{$\;\;\;\,\to\;\;\;\,$}}

\def\colim{\raisebox{-1.2ex}{\rlap{\tiny $\;\;\;p_i \parallel p_j$}}
\raisebox{0ex}{$\;\;\;\,\to\;\;\;\,$}}

\def\cflim{\raisebox{-1.2ex}{\rlap{\tiny $\;\;\;p_i \parallel p_\gamma$}}
\raisebox{0ex}{$\;\;\;\,\to\;\;\;\,$}}

\def\icflim{\raisebox{-1.2ex}{\rlap{\tiny $\;p_i \parallel p_A$}}
\raisebox{0ex}{$\;\;\;\,\to\;\;\;\,$}}

\def\cflimn{\raisebox{-1.2ex}{\rlap{\tiny $\;p_{n+1} \parallel p_\gamma$}}
\raisebox{0ex}{$\;\;\;\,\to\;\;\;\,$}}

\mysection{Isolated-particle production and factorization}\label{factorisation}

\noindent
When no isolation criterion is applied, the inclusive photon 
cross section is computable by using the QCD factorization formula
(\ref{2.1}). The formula states that at large values of $p_{ T \, \gamma}$
the short-distance dynamics is perturbatively computable in terms of partonic 
cross sections, while the dominant non-perturbative phenomena can be factorized
in the parton densities of the colliding hadrons and the fragmentation
function of the detected photon. Owing to the inclusiveness of the process,
all the remaining non-perturbative effects
have the form $(Q_0/p_{ T \, \gamma})^p$ and are suppressed by some inverse 
power $p$ of the photon transverse momentum. These terms are not indicated on 
the right-hand side of eq.~(\ref{2.1}), since they are negligible as long as
$p_{ T \, \gamma}$ is much larger than the typical hadronic scale 
$Q_0 \sim {\cal O}(1$~GeV).

\vspace{0.3cm}

\noindent
In the case of isolated photons, the isolation criterion 
(\ref{criterion}) enforces additional phase-space restrictions. 
This implies that the cross section is {\em no} longer fully inclusive over the
hadronic final state and, hence, that
the factorized expression (\ref{2.1}) is not necessarily valid
\cite{berger-qiu, berger-guo-qiu, Aurenche:1997ng, cfp}. This 
issue was considered in ref.~\cite{cfp}, where factorization was proved
at {\em any} order in QCD perturbation theory in the case of the isolation
criterion of eq.~(\ref{2.1}). The discussion in ref.~\cite{cfp} was mainly
focused on prompt-photon production in $e^{+}e^{-}$ collisions.
To better clarify the differences between $e^{+}e^{-}$ collisions and
hadron--hadron collisions,
in the following we reconsider in more detail the latter case.

\vspace{0.3cm}
\noindent
In the theoretical literature 
(see e.g. refs.~\cite{berger-qiu,gordon-vogelsang2}) on isolated-photon
production at hadron colliders, the isolation criterion (\ref{criterion})
is sometime replaced by a similar one, in which the isolation is defined 
by an upper limit $E_{max}$ on
the accompanying {\em hadronic energy} $E^{had}$ in the 
c.m. frame rather than 
by an upper limit $E_{T\, max}$ on
the hadronic transverse energy
$E_T^{had}$. To discuss both criteria in a unified manner, we introduce
a more exclusive definition of the isolation. The photon is isolated if the total
hadronic four-momentum $Q^{had}_\mu$
inside the isolation cone is {\em fixed} at a given value $Q_{\mu}$:
\begin{equation}\label{gencriterion}     
Q^{had}_\mu =  Q_{\mu}\;\;\;\; \mbox{inside} \;\;\;\;      
\left( y - y_{\gamma} \right)^{2} +    
\left(  \phi - \phi_{\gamma} \right)^{2}  \leq R^{2}  \;.    
\end{equation}     
The cross section corresponding to this generalized isolation is denoted by
$d\sigma^{g-is}/d^4Q$. From it, we can recover
the cross section $\sigma^{is}$ corresponding to the criterion 
in eq.~(\ref{criterion}) by simply integrating over $Q_{\mu}$ as
\begin{equation}\label{genxs}
\sigma^{is} = \int d^4Q \; \frac{d\sigma^{g-is}}{d^4Q}
\; \Theta\!\left( E_{T \, max} - E_{T} \right) \;,
\end{equation}   
where $E_{T}$ denotes the transverse energy of the four-momentum $Q_\mu$.
We can also replace the hadronic transverse energy with the hadronic energy
in the isolation criterion (\ref{criterion}), by simply replacing the transverse
energies $E_{T \, max}, E_{T}$ with the corresponding c.m. energies
$E_{max}, E$ in the argument of the $\Theta$-function on the right-hand side
of eq.~(\ref{genxs}).
In the following we discuss factorization for the generalized isolated cross
section $d\sigma^{g-is}/d^4Q$.

\vspace{0.3cm}

\noindent
Our proof of factorization is based on the consistency, at the level of
{\em power unsuppressed} contributions, between QCD perturbation theory and
the full QCD theory. We exploit the formal correspondence between singularities 
in the calculation at the parton level and non-perturbative effects.
Since the hadronic cross section is computable and finite in the full theory,
the presence of singularities that are not cancelled or not factorizable
at the parton level implies that of compensating singularities of 
non-perturbative origin. Thus, we are led to consider the perturbative 
singular behaviour, which is due to the emission of soft and collinear partons.
This picture obviously agrees with the physical expectation that 
factorization can only be spoiled by long-distance phenomena, such as those
involving soft and collinear radiation. 

\vspace{0.3cm}

\noindent
To give our formal proof, we consider the production of the prompt photon
with momentum $p_{\gamma}$ in the collisions of two hadrons $H_A$ and $H_B$ with 
momenta $p_A$ and $p_B$. In general, the hadronic final state contains $n+1$ 
(with $n \geq 0$, since the transverse momentum of the photon has to be balanced
by some final-state radiation) additional partons (hadrons) with momenta 
$p_1, \dots,p_{n+1}$. Any isolation criterion applied to the photon
is thus specified in terms of a function 
\begin{equation}
\label{fn} 
F_{\{is\}}^{(n+1)}(p_A,p_B;p_\gamma;p_1, \dots p_{n+1}) 
\end{equation} 
that depends on the momenta of the particles in the event. The subscript
$\{is\}$ denotes the dependence on the isolation parameters, whose
precise definition is not explicitly spelled out for the moment.
The factorization of the prompt-photon cross section with isolation
can be studied \cite{slice, submeth, dipmeth}
in terms of the properties of the isolation function in
eq.~(\ref{fn}). More precisely, factorization is valid provided 
$F_{\{ is\}}$ fulfils the following requirements\footnote{In 
eqs.~(\ref{irsafe})--(\ref{icfact}) bold-face characters are used
just to emphasize the differences between left-hand and right-hand sides.}

\vspace{0.3cm}
\noindent $i)$ infrared safety:
\begin{equation}
\label{irsafe}
F_{\{ is\}}^{(n{\bom {+1}})}(p_A,p_B;p_\gamma;p_1, \dots {\bom p_i}, 
\dots p_{n+1}) 
\irlim
F_{\{ is\}}^{(n)}(p_A,p_B;p_\gamma;p_1, \dots p_{n+1}) \;,
\end{equation}

\noindent $ii)$ collinear safety:
\begin{equation}
\label{csafe}
F_{\{ is\}}^{(n{\bom {+1}})}(p_A,p_B;p_\gamma;p_1, \dots {\bom p_i}, 
{\bom p_j}, \dots p_{n+1}) 
\colim 
F_{\{ is\}}^{(n)}(p_A,p_B;p_\gamma;p_1, \dots, {\bom {p_i + p_j}}, \dots p_{n+1}) \;,
\end{equation}

\noindent $iii)$ final-state collinear factorizability:
\begin{equation}\label{cfact}
F_{\{ is\}}^{(n{\bom {+1}})}(p_A,p_B;{\bom p_\gamma};p_1, \dots {\bom p_i}, \dots 
p_{n+1}) 
\cflim
F_{\{ is\}}^{(n)}(p_A,p_B;{\bom {p_\gamma+p_i}};p_1, \dots \dots p_{n+1}) 
\;,
\end{equation}

\noindent $iv)$ initial-state collinear factorizability:
\begin{equation}
\label{icfact}
F_{\{ is\}}^{(n{\bom {+1}})}({\bom p_A},p_B;p_\gamma;p_1, \dots {\bom p_i}, \dots 
p_{n+1}) 
\icflim
F_{\{ is\}}^{(n)}({\bom {p_A - p_i}},p_B;p_\gamma;p_1, \dots \dots p_{n+1}) 
\;.
\end{equation}
Of course, the analogue of eq.~(\ref{icfact}) with $p_A \leftrightarrow p_B$
is understood.

\vspace{0.3cm}

\noindent
The requirement of $i)$ infrared safety 
means that the cross section is insensitive to the momenta of arbitrarily 
soft particles. The requirement of $ii)$ collinear safety implies that,
when some final-state particles are produced collinearly, the 
cross section depends on their total momentum rather than on the momentum of 
each of them. The general isolation criterion in eq.~(\ref{gencriterion}) 
depends on $Q_{\mu}= Q_{\mu}^{had} = \sum_{j \in { cone}} p_{\mu \, j}$,
the total hadronic momentum inside the isolation cone.
Since $Q_{\mu}^{had}$ is an infrared- and collinear-safe quantity, the 
isolation criterion in eq.~(\ref{gencriterion}) fulfils eqs.~(\ref{irsafe}) and 
(\ref{csafe}).

\vspace{0.3cm}

\noindent
The property in eq.~(\ref{cfact}) guarantees that all the long-distance
phenomena related to the low-momentum fragmentation of the photon can be 
absorbed and factorized in the universal fragmentation
function $D_a^\gamma(z; M_F)$. By universal we mean that it does
not depend on the process and, in particular, it does not depend
on the isolation parameters. Naive inspection of the isolation criterion
in eq.~(\ref{gencriterion}) may suggest that it violates eq.~(\ref{cfact}).
As pointed out in ref.~\cite{cfp}, this is not the case.

\vspace{0.3cm}

\noindent
To explain the key point \cite{cfp}, let us first consider the following 
isolation criterion:
\begin{equation}
\label{gfright} 
F_{\{Q^{ cut},R\}}^{(n+1)}(p_A,p_B;p_\gamma;p_1, \dots p_{n+1}) =
\delta^{(4)}\!\left( Q^{ cut} - 
\left[ p_{\gamma} +
\sum_{j=1}^{n+1} p_{j} \; \Theta(R - R_{j \gamma}) \right] \right) \;\;,
\end{equation} 
where
\begin{equation}
\label{angdis}
R_{j \gamma} =  \sqrt {(y_j -y_\gamma)^2 + (\phi_j - \phi_\gamma)^2} \;\;,
\end{equation}
and $R$ and 
$Q_\mu^{ cut}$ are {\em external} isolation parameters.
The term in the square bracket on the right-hand side of 
eq.~(\ref{gfright}) is the total (hadron+photon) 
four-momentum inside the 
cone of radius $R$. Thus we have
\begin{equation}
\label{getcut}
p_{\gamma} + \sum_{j=1}^{n+1} p_{j} \; \Theta(R - R_{j \gamma}) 
\cflim
( p_{\gamma}+ p_{i} )+ \sum_{\stackrel{j=1}{j\neq i}}^{n+1} 
p_{j} \; \Theta(R - R_{j \gamma}) \;\;,
\end{equation}
and the function 
$F_{\{Q^{cut},R\}}^{(n+1)}$ 
fulfils eq.~(\ref{cfact}).

\vspace{0.3cm}

\noindent
Then, we can observe that the momentum $p_\gamma$
of the photon and the isolation parameter 
$Q_{\mu}$ in eq.~(\ref{gencriterion})
are both kept fixed in the measurement of the cross section, so they can be 
regarded as {\em external} variables that are independent of the momenta 
$p_A,p_B,p_\gamma,p_1, \dots p_{n+1}$. Therefore, by simply making the 
identification 
\begin{equation}
\label{qcutvsq}
Q_\mu^{cut} = Q_{\mu} + p_{\mu \,\gamma} \;,
\end{equation} 
the isolation
criterion in eq.~(\ref{gencriterion}) can be recast in the form of 
eq.~(\ref{gfright}), which manifestly satisfies eq.~(\ref{cfact}).

\vspace{0.3cm}

\noindent
The properties in eqs.~(\ref{irsafe})--(\ref{cfact}) are sufficient to prove
factorization in the case of $e^+e^-$ collisions \cite{cfp}. In hadron--hadron
collisions, the cross section is affected by additional long-distance phenomena
related to the non-perturbative binding of the colliding partons into the
incoming hadrons. At the parton level, these phenomena lead to
initial-state collinear singularities that have to be absorbed and factorized
in the non-perturbative parton distributions of the hadrons $H_A$ and $H_B$.
The property in eq.~(\ref{icfact}) guarantees that
the photon-isolation criterion does not spoil 
the factorization of the initial-state collinear singularities.
Since the expression on the right-hand side of eq.~(\ref{gfright}) does not
explicitly depend either on the incoming momenta $p_A,p_B$ or on any 
final-state momentum parallel to them, the property in eq.~(\ref{icfact})
is thus evidently fulfilled by the generalized isolation criterion in 
eq.~(\ref{gencriterion}).

\vspace{0.3cm}

\noindent
The main conclusion of our discussion on the generalized isolation criterion in 
eq.~(\ref{gencriterion}) is that QCD factorization is valid at {\em any} order
in perturbation theory. Factorization for the isolation criterion  
(\ref{criterion}) with respect to the hadronic transverse energy (or to the
hadronic energy) thus follows from eq.~(\ref{genxs}). Note, however,
that $d\sigma^{g-is}/d^4Q$ factorizes as a function of the fixed isolation
parameter $Q_\mu^{cut}$ rather than as a function of $Q_\mu$ (see
eqs.~(\ref{gfright}) and (\ref{qcutvsq})). As discussed below,
this functional dependence has
influence upon the kinematical structure of the factorization formula
for the isolated cross sections.

\vspace{0.3cm}
\noindent
We first discuss the isolation criterion (\ref{criterion}) with respect to
the hadronic transverse energy. Since $d\sigma^{g-is}/d^4Q$ factorizes at
fixed $Q_\mu^{cut}=Q_\mu+p_{\mu \,\gamma}$, the constrained integration
in eq.~(\ref{genxs}) leads to a dependence on the variable $E_T^{cut}$:
\begin{equation}
\label{Etcut}
E_T^{cut} = E_{T \, max} + E_{T \,\gamma} = E_{T \, max} + p_{T \,\gamma}\;.
\end{equation} 
The inclusive cross section is thus a function 
on the photon momentum $p_{\gamma}$ and on the isolation
parameters $R$ and $E_T^{cut}$.
It is convenient to define the 
variable
\begin{equation}
\label{zcut}
z_c \equiv \frac{p_{T \, \gamma}}{E_T^{cut}} 
= \frac{p_{T \, \gamma}}{E_{T \, max} + p_{T \, \gamma}} < 1\;\;.
\end{equation}
The inclusive distribution
$d\sigma^{{is}}/dp_{ T \, \gamma}dy_{\gamma}$ with transverse-energy isolation,
which we simply denote by $\sigma^{{is}}(p_{\gamma};z_c,R)$,
fulfils a factorization formula analogous to eq.~(\ref{2.1}):
\begin{eqnarray}
\label{isxsp} 
\!\!\!\!\!\!\!\!\!\!\!\!\!\!\!
\sigma^{{is}}(p_{\gamma};z_c,R) \!\!\!\!\!&=&\!\!\!\!\!
\sum_{a} \int_0^1 \frac{dz}{z} \;
\widehat{\sigma}^{a,{is}}\left(\frac{p_{\gamma}}{z};
\frac{z_c}{z},R;\mu,M,M_{F}\right)  
D_{a}^{\gamma}(z;M_F)
+ \widehat{\sigma}^{\gamma,{is}}(p_{\gamma};z_c,R;\mu,M,M_{F}) \\
\label{isxs}
\!\!\!\!\!&=&\!\!\!\!\! \sum_{a} \int_0^1 \frac{dz}{z} \;
\widehat{\sigma}^{a,{is}}\left(\frac{p_{\gamma}}{z};
\frac{z_c}{z},R;\mu,M,M_{F}\right) 
D_{a}^{\gamma}(z;M_F) \;\Theta(z-z_c)  \\
\!\!\!\!\!&+&\!\! 
\widehat{\sigma}^{\gamma,{is}}(p_{\gamma};z_c,R;\mu,M,M_{F}) \;.\nonumber
\end{eqnarray}
The fragmentation function $D_{a}^{\gamma}(z;M_F)$ is the same
fragmentation function as appears in the non-isolated case. In particular,
it does not depend on the isolation parameters. The dependence on the latter
is fully embodied in the 
subprocess cross sections
$\widehat{\sigma}^{a,{is}}$ and $\widehat{\sigma}^{\gamma,{is}}$,
which respectively give the fragmentation and direct contributions to
the hadronic cross section. We recall that the subprocess cross sections
$\widehat{\sigma}^{a,{is}}$ and $\widehat{\sigma}^{\gamma,{is}}$
are obtained by convoluting the parton densities of the colliding hadrons
with the cross sections 
$\widehat{\sigma}_{ij}^{a,{is}}$ and 
$\widehat{\sigma}_{ij}^{\gamma,{is}}$ of the partonic subprocesses
$i+j \to a+X$ and $i+j \to \gamma+X$. We have
\begin{eqnarray}
\widehat{\sigma}^{a,{is}}\left(\frac{p_{\gamma}}{z};
\frac{z_c}{z},R;\mu,M,M_{F}\right) &=& \sum_{i,j} \int_0^1 dx_1 \;dx_2 \;
G_{i/A}(x_1,M) \; G_{j/B}(x_2,M) \nonumber \\
\label{sigpartonic}
&\times& \widehat{\sigma}_{ij}^{a,{is}}\left(x_1p_A,x_2p_B, \frac{p_{\gamma}}{z};
\frac{z_c}{z},R;\mu,M,M_{F}\right) \;,
\end{eqnarray}
and a similar formula relates $\widehat{\sigma}^{\gamma,{is}}$ to
$\widehat{\sigma}_{ij}^{\gamma,{is}}$.

\vspace{0.3cm}

\noindent
Note that, according to eqs.~(\ref{gfright}) and (\ref{genxs}), 
factorization holds at fixed
$E_T^{cut}$. Since $z_c$ is obtained by rescaling $E_T^{cut}$ with
the factor $p_{ T \, \gamma}$, the variable $z_c$ is a {\em scaling} variable
with respect to factorization. In other words, the partonic cross section
in eq.~(\ref{isxs}) depends on $p_\gamma/z$ and $z_c/z$. In particular,
since $z_c$ is constrained to be $z_c < 1$ from eqs.~(\ref{gfright}) 
and (\ref{genxs}),
this constraint propagates to 
$\widehat{\sigma}^{a,{is}}(p_\gamma/z;z_c/z)$ as $z_c/z < 1$. 
We made this condition explicit in eq.~(\ref{isxs}).

\vspace{0.3cm}

\noindent
In current experimental practice, $E_{T \, max}$ is sometimes
expressed
in terms of the dimensionless parameter $\varepsilon_{h}$ defined by
\begin{equation}
\label{epsilon-h}
\varepsilon_{h} = \frac{E_{T\, max}}{p_{T \, \gamma}} \;\;.
\end{equation} 
This parameter is related to our scaling variable $z_c$ by
\begin{equation}
z_{c} = \frac{1}{1 + \varepsilon_{h}} \;\;.
\end{equation}

\vspace{0.3cm}

\noindent
The partonic cross sections in eq.~(\ref{isxs}) can be expanded
as power series in $\alpha_s$ analogously to the fully inclusive case in 
eqs.~(\ref{isa}) and (\ref{2.2b}). Actually, at the Born level, 
it is straightforward to show \cite{berger-qiu} 
that $\widehat{\sigma}^{\gamma,{is}}$
and $\widehat{\sigma}^{a,{is}}$ exactly coincide with the corresponding
expression for the non-isolated case, apart from the overall constraint
$z_c/z < 1$ mentioned above.
Up to NLO, we thus have
\begin{eqnarray}
\!\!\!\!\!\!\!\!\!\!\!\!
\widehat{\sigma}^{\gamma,{is}}(p;z_c,R;\mu ,M,M_{F}) \!\!\!\!\!&=&\!\!\!\!\! 
\left ( \frac{\alpha_s(\mu )}{\pi} \right ) \;
\sigma_{Born}^{\gamma}(p;M) + 
\left( \frac{\alpha_s(\mu)}{\pi} \right)^{2} \!
\sigma_{HO}^{\gamma,{is}}(p;z_c,R;\mu,M, M_{F}) \;, \label{isa} \\
\!\!\!\!\!\!\!\!\!\!\!\!
\widehat{\sigma}^{a,{is}}(p;z_c,R;\mu,M,M_{F}) \!\!\!\!\!&=&\!\!\!\!\! 
\left( \frac{\alpha_s(\mu)}{\pi} \right)^{2} \!\!
\sigma^{a}_{Born}(p;M) + 
\left( \frac{\alpha_s(\mu)}{\pi} \right)^{3} \!
\sigma^{a,{is}}_{HO}(p;z_c,R;\mu,M,M_F) \;. \label{isb}
\end{eqnarray} 
In the following sections we compute the NLO terms 
$\sigma_{HO}^{\gamma,{is}}$ and $\sigma^{a,{is}}_{HO}$.

\vspace{0.3cm}

\noindent
We can now consider the variant of the isolation criterion (\ref{criterion})
with respect to the hadronic c.m. energy. 
We can straightforwardly follow the previous discussion
on transverse-energy isolation. The transverse-energy
isolation parameter $E_T^{cut}$
in eq.~(\ref{Etcut}) and the scaling variable $z_c$ in eq.~(\ref{zcut})
have to be respectively replaced by the energy isolation parameter
$E^{cut}$ and the scaling variable $z_c^E$:
\begin{equation}
E^{cut} = E_{max} + E_{\gamma} \;,
\end{equation} 
\begin{equation}
z_c^E \equiv \frac{E_{\gamma}}{E^{cut}} 
= \frac{E_{\gamma}}{E_{max} + E_{\gamma}} < 1\;\;.
\end{equation}
Denoting by $\sigma^{{E-is}}$ the energy-isolation variant of the inclusive
distribution $\sigma^{{is}}$, we still have a factorization formula
analogous to eq.~(\ref{isxs}): it is sufficient to make the formal
replacements
\begin{eqnarray}
\sigma^{{is}}(p_{\gamma};z_c,R) & \to &
\sigma^{{E-is}}(p_{\gamma};z_c^E,R) \;, \nonumber \\
\label{isvseis}
\widehat{\sigma}^{{is}}\left(\frac{p_{\gamma}}{z};
\frac{z_c}{z},R;\mu,M,M_{F}\right) & \to &
\widehat{\sigma}^{{E-is}}\left(\frac{p_{\gamma}}{z};
\frac{z_c^E}{z},R;\mu,M,M_{F}\right) \;, \\
z_c & \to & z_c^E \;. \nonumber 
\end{eqnarray}
In particular, the fragmentation function $D_{a}^{\gamma}(z;M_F)$ is not
affected by the change of the isolation criterion.
Note, however, that the strict formal correspondence between the two variants of 
the isolation criterion does not extend to the cross sections of the partonic
subprocesses $i+j \to a+X$ and $i+j \to \gamma+X$.
In fact, although the convolution structure of eq.~(\ref{sigpartonic}) is still
valid, we have to perform the following replacement:
\begin{equation}
\label{reppart}
\widehat{\sigma}_{ij}^{a,{is}}\!\!\left(x_1p_A,x_2p_B, \frac{p_{\gamma}}{z};
\frac{z_c}{z},R;\mu,M,M_{F}\right)
\to
\widehat{\sigma}_{ij}^{a,{E-is}}\!\!\left(x_1p_A,x_2p_B, \frac{p_{\gamma}}{z};
\frac{z_c^E}{z},R,
\frac{E^{cut}}{x_1\sqrt S},\frac{E^{cut}}{x_2\sqrt S};
\mu,M,M_{F}\right) \!.
\end{equation}
Unlike $\widehat{\sigma}_{ij}^{a,{is}}$, its energy-isolation variant
$\widehat{\sigma}_{ij}^{a,{E-is}}$ does depend on 
$E^{cut}/x_1\sqrt S$ and $E^{cut}/x_2\sqrt S$. This additional
dependence follows from having defined isolation with respect to the hadronic
energy in the c.m. frame. Such a definition is not invariant
under longitudinal boosts along the beam direction, thus leading to an entangled
dependence on $E^{cut}$ and on the energies $x_1{\sqrt S}/2, 
x_2{\sqrt S}/2$ of the partonic beams. In other words, the dependence on the
isolation parameter $E^{cut}$ is not {\em kinematically} factorized
\cite{cfp} from the dependence on the momentum fractions in the parton 
densities.

\vspace{0.3cm}
\noindent
Despite the formal correspondence in eq.~(\ref{isvseis}), 
beyond the LO, the functional dependence of
$\sigma^{{E-is}}$ on $z_c^E$ and $R$ is not the same as 
the functional dependence of $\sigma^{{is}}$ on $z_c$ and $R$.
However, the transverse-energy 
fraction $\sum_j E_{T \, j}/E_{T \, \gamma}$ and the 
c.m.-energy fraction $\sum_j E_{j}/E_{\gamma}$ coincide
as long as all the partons (hadrons) $j$ inside the isolation cone
are either soft or collinear to the photon direction. This implies
that the two variants of the isolation criterion can {\em substantially} 
differ only if the isolation cone contains at least one {\em hard} parton 
that is not collinear to the photon. Such a kinematical 
configuration is suppressed
both in the (soft) limit $z_c \to 1$ and in the (collinear) limit $R \to 0$.
In either of these limits, the two variants of the isolation criterion
perturbatively coincide: the order-by-order perturbative calculations of
$\sigma^{{is}}$ and $\sigma^{{E-is}}$ differ by terms that are of
${\cal O}(1-z_c)$, when $z_c \to 1$, and of ${\cal O}(R^2)$, when $R \to 0$.

\vspace{0.3cm}

\noindent
The validity of factorization implies that the partonic cross sections in
eqs.~(\ref{isa}) and (\ref{isb}) are computable in QCD perturbation theory. 
Nonetheless, their fixed-order perturbative expansions are not always
well behaved. This is the case, for instance, in the kinematical configurations
of highly isolated photons ($1-z_c \ll 1$) and of very small isolation cones
($R \ll 1$). When $1-z_c \ll 1$, the fixed-order expansion contains large
double-logarithmic contributions, $(\alpha_s\ln^2 (1-z_c))^n$, of soft origin.
When $R \ll 1$, the fixed-order expansion contains large
single-logarithmic contributions, $(\alpha_s\ln R)^n$, of collinear origin.
The effects of these logarithmic contributions at NLO are discussed in 
Sect.~\ref{numcalc}.

\vspace{0.3cm}

\noindent
Another general source of misbehaviour in the fixed-order expansion
of perturbatively computable observables is the possible presence of integrable
logarithmic singularities at some `critical' points away from the soft and 
collinear boundaries of the phase space \cite{catani-webber}. Such 
singularities occur \cite{berger-guo-qiu,Aurenche:1997ng,cfp}
in the spectrum of isolated prompt photons produced in
$e^+e^-$ collisions, and are located at the critical point
$2E_\gamma=z_c^E {\sqrt S}$ (in $e^+e^-$ collisions, the isolation
is defined with respect to the c.m. energies).
The partonic cross section
${\sigma}_{ij}^{a,{is}}$ (or ${\sigma}_{ij}^{a,{E-is}}$)
in eq.~(\ref{reppart}) 
has an analogous critical point at
$2E_{T \,\gamma}=z_c {\sqrt {x_1x_2S}}/\cosh y_{\gamma}^*$
(or $2E_{T \,\gamma}=z_c^E {\sqrt {x_1x_2S}}/\cosh y_{\gamma}^*$),
where $y_{\gamma}^*= y_{\gamma} - \ln {\sqrt {x_1/x_2}}$ is the photon
rapidity in the partonic c.m. frame.
Unlike the c.m. energy ${\sqrt S}$ in $e^+e^-$ collisions, 
the partonic c.m. energy $\sqrt{\hat{s}}={\sqrt {x_1x_2S}}$
is not fixed and depends on the momentum fractions $x_1$ and $x_2$ of the
colliding partons. Any possible {\em integrable} singularities in the
partonic cross sections ${\sigma}_{ij}^{a,{is}}$ thus disappear
after integration over $x_1$ and $x_2$ (see eq.~(\ref{sigpartonic})).
In conclusion, there are no critical points in the single-photon
inclusive cross sections $\sigma^{{is}}(p_{\gamma};z_c,R)$, 
$\widehat{\sigma}^{\gamma,{is}}(p;z_c,R;\mu ,M,M_{F})$ and
$\widehat{\sigma}^{a,{is}}(p;z_c,R;\mu ,M,M_{F})$
(see eqs.~(\ref{isxs}), (\ref{isa}) and (\ref{isb})) in which we are interested 
throughout the present paper. Note, however, that critical points and related
integrable singularities occur in less inclusive distributions of isolated
prompt photons produced in hadron--hadron collisions
(see e.g. ref.~\cite{bgpw}).

\vspace{0.3cm}

\noindent
The photon isolation procedure actually implemented by the D0 Collaboration
\cite{d0} does not exactly coincide with the isolation criterion in 
eq.~(\ref{criterion}). A cone of radius $R$ around the photon direction
is still considered, but the upper limit $E_{T \, max}$ is enforced on the
hadronic transverse energy inside an annular region of width $\Delta R$
($\Delta R < R$), rather than on the hadronic transverse energy inside
the whole cone. More precisely, eq.~(\ref{criterion}) is replaced by the
following:
\begin{equation}\label{d0criterion}     
E_{T}^{had}(R) -  E_{T}^{had}(R-\Delta R) \leq E_{T \, max} \;, 
\end{equation}  
where $E_{T}^{had}(R)$ and $E_{T}^{had}(R-\Delta R)$ are the hadronic 
transverse energies in the two cones of radius $R$ and $R-\Delta R$,
respectively. This criterion is thus specified by the isolation function
\begin{equation}
\label{d0fright} 
F_{\{E_{T \, max},R,\Delta R\}}^{(n+1)}(p_A,p_B;p_\gamma;p_1, \dots p_{n+1}) =
\Theta\!\!\left( E_{T \, max} - 
\sum_{j=1}^{n+1} E_{T \,j} \; \Theta(R - R_{j \gamma}) \;
\Theta(R_{j \gamma} - R + \Delta R ) \right) \;\;.
\end{equation} 
It is straightforward to check that the properties in 
eqs.~(\ref{irsafe})--(\ref{icfact}) are fulfilled by eq.~(\ref{d0fright}),
so the criterion (\ref{d0criterion}) fulfils factorization. However, the
kinematical structure of the corresponding factorized cross section,
denoted by $\sigma^{{D0-is}}(p_{\gamma};E_{T \, max},R,\Delta R)$, 
is different from that in eq.~(\ref{isxs}). We have:
\begin{eqnarray}
\!\!\!\!\!\!\!\!\!\!\!\!
\sigma^{{D0-is}}(p_{\gamma};E_{T \, max},R,\Delta R) 
\!\!\!\!\!&=&\!\!\!\!\! \sum_{a} \int_0^1 \frac{dz}{z} \;
\widehat{\sigma}^{a,{D0-is}}\left(\frac{p_{\gamma}}{z};
E_{T \, max},R,\Delta R;\mu,M,M_{F}\right) 
D_{a}^{\gamma}(z;M_F) 
\nonumber \\
\label{d0isxs}
\!\!\!\!\!&+&\!\! 
\widehat{\sigma}^{\gamma,{D0-is}}(p_{\gamma};E_{T \, max},
R,\Delta R;\mu,M,M_{F}) \;.
\end{eqnarray}
The differences between eqs.~(\ref{isxs}) and (\ref{d0isxs}) arise from the fact
that the criterion (\ref{d0criterion}) does not constrain the hadronic
transverse energy collinear to the photon. Thus the isolation parameter
$E_{T \, max}$ is not rescaled by the photon momentum fraction $z$ when going
from the hadronic cross section $\sigma^{{D0-is}}$ on the left-hand side of
eq.~(\ref{d0isxs}) to the partonic cross
section $\widehat{\sigma}^{a,{D0-is}}$ 
on the right-hand side. Correspondingly,
the isolation procedure does not set any absolute lower limit (such as
$z > z_c$ in eq.~(\ref{isxs})) on the momentum fraction $z$ in the photon
fragmentation function. In particular, at the LO the isolated cross section
$\sigma^{{D0-is}}$ exactly coincides with the non-isolated cross section in
eq.~(\ref{2.1}). Of course, higher-order contributions to the 
isolated cross section are different from those to the non-isolated one,
and tend to suppress the direct and fragmentation components.

\vspace{0.3cm}

\noindent
We conclude this section with two general observations. We have explicitly 
discussed isolated-photon production only in the case of hadron--hadron
collisions. However, our discussion straightforwardly applies also to
photon--hadron and photon--photon collisions: it is sufficient to substitute 
the parton densities of the colliding hadron for those of the colliding photon.
Analogously, the final-state isolated photon can be replaced by any 
final-state isolated hadron (e.g. a pion) by substituting the fragmentation
functions of the photon for those of the hadron.

\mysection{Isolated cross section}\label{ancalc_isol}

After the all-order proof of factorization,  
we explicitly study how the isolation criterion (\ref{criterion})
modifies the inclusive cross section calculated at NLO. In particular,
we show how the partonic cross sections $\sigma^{\gamma,{\rm is}}_{HO}$
and $\sigma^{a,{\rm is}}_{HO}$ depend on the cone size $R$. We also show
how the constraint on the transverse energy inside the isolation cone 
can be translated into conditions on the integration range
of the longitudinal variables in eqs.~(\ref{2.7}), (\ref{2.8}) and (\ref{2.13}).
In our calculation we follow a procedure similar to that in 
refs.~\cite{owens,gordon-vogelsang2}: we start from the inclusive cross section
and `subtract' the contributions that do not fulfil the isolation criterion.

\subsection{Direct contribution with isolation}\label{dirwis}

\noindent
Considering the ${\cal O}(\alpha \alpha_{s}^{2})$ contribution to the direct 
component (see fig.~\ref{Fig:kinematics}), when the momentum $\vec{p}_{q}$ of the final-state quark
is inside the cone around the photon,
the isolation criterion (\ref{criterion})
can be written as

\[
p_{T \, q} \leq \varepsilon_{h} \ p_{T \, \gamma} \;,
\]
or, equivalently,
\begin{equation}\label{3.1}
\frac{p_{T \, \gamma}}{p_{T \, q} + p_{T \, \gamma}} \geq 
\frac{1}{1 + \varepsilon_{h}} \equiv z_{c} \;.
\end{equation} 

\noindent 
Therefore, the contribution to the isolated cross section is obtained from
that to the inclusive cross section by subtracting from the latter 
the part that violates the constraint (\ref{3.1}).

\vspace{0.3cm}

\noindent
In terms of the result in eq.~(\ref{2.18}), obtained within the collinear approximation,
eq.~(\ref{3.1}) is 
implemented by the condition $z > z_{c}$. Therefore the direct
contribution to the isolated cross section is obtained from eq.~(\ref{2.18})
by simply subtracting the term

\begin{eqnarray}
\label{3.2}
d\sigma_{\gamma jet}^{direct,sub} & = &
2 \pi \left( \frac{\alpha_{s} (\mu)}{\pi} \right)^{2} 
\sum_{b\atop{a= q,\bar{q}}} \int_{z_{min}}^{z_{c}} E_{a} 
\frac{d \widehat{\sigma}^{Born}({\scriptstyle A + B \to a + b})}
     {d \vec{p}_{a} \ d\eta_{b}} \ 
d\eta_{b} \ d\eta_{a} \ p_{ T \, a }^{n-3} \ dp_{ T \, a } \nonumber \\
& &
\;\;\;\;\;\;\;\;\;\;\;\;\;\;\;\;\;\;\;\;\; \times \;\;
\frac{\alpha}{2 \pi} \ e_{a}^{2} 
\left[ 2 \ln \left( \frac{R \ p_{T \, \gamma} \ (1-z)}{M_{F}} \right) 
\frac{1 + (1-z)^{2}}{z} + z \right] dz \;\;.
\end{eqnarray}

\noindent
We recall that expression (\ref{3.2}) has 
been obtained in the approximation $R \ll 1$. All the terms of  
${\cal O}(R^{2n})$ ($n \geq 1$) have been neglected and will be calculated as 
discussed in sect.~\ref{numcalc}. 

\vspace{0.3cm}

\noindent
Note that the particular choice $M_{F} = R \ p_{T\gamma}$ eliminates the
$R$-dependence from eq.~(\ref{3.2}). The price to pay is the introduction of 
an additional 
$R$-dependence in other terms of the cross section, notably through the
$M_{F}$-dependence of the fragmentation function.

\vspace{0.3cm}

\noindent
Note also that when the momentum $p_{q}$ of the final-state quark in 
fig.~\ref{Fig:kinematics} is outside the isolation cone,
the isolation criterion (\ref{criterion}) does not enforce any additional
constraint. In particular, the HO contribution in eq.~(\ref{outside}) is left
unchanged in the isolated case. As mentioned at the end of 
sect.~\ref{ancalc_incl}, the 
$\ln R$-dependence of eq.~(\ref{outside}) exactly cancels the 
$\ln R$-dependence of eq.~(\ref{2.18}). Since in the isolated case, we have to
add eqs.~(\ref{2.18}) and (\ref{outside}) and then subtract eq.~(\ref{3.2}),
the cancellation of the $\ln R$-dependence does not occur anymore.
When $R \ll 1$, the NLO calculation of the isolated cross section is
proportional to $\ln 1/R$, and therefore it diverges to $+\infty$ when $R \to
0$. 

\vspace{0.3cm}

\noindent
This divergence is unphysical. Its appearance at NLO simply means that, as soon
as $R$ is sufficiently small (see sect.~\ref{numcalc}), 
the NLO calculation is not physically reliable.
To improve the 
reliability of the fixed-order
perturbative expansion, higher-order
contributions proportional to $(\alpha_s \ln R)^n$ have to be computed when 
$R$ becomes very small.

\vspace{0.3cm}

\noindent
Moreover, the calculation in sect.~\ref{ancalc_incl} also tells us that non-perturbative
contributions must be taken into account when $R$ is very small. For example,
the integral in eq.~(\ref{2.14}) is an integral over the virtuality 
$p_{a}^{2}$ of the parton $a$ in fig.~\ref{Fig:kinematics}. The integral 
of the perturbative matrix element is
performed over the range $0 \leq p_{a}^{2} = 2 p_{q}.p_{\gamma} \leq R^{2} \, p_{T
\,\gamma}^{2}  \, (1-z)/z$.
However, non-perturbative hadronization effects become dominant as soon as
$p_{a}^{2} \leq Q_{0}^{2} \sim {\cal O}(1$~GeV$^{2})$.
A strict perturbative treatment of the isolated cross section is thus
justified only when
\begin{equation}\label{cond-valid}
R^{2} \, p_{T \, \gamma}^{2} \, \frac{1-z}{z} \geq Q_{0}^{2} \;\;.
\end{equation}
If $R$ becomes too small, the perturbative calculation has to be supplemented
by a careful treatment of non-perturbative phenomena.

\subsection{Fragmentation contribution with isolation}\label{isol-fragm}

\noindent
To study the effects of isolation on the fragmentation component, let us consider the
subprocess of fig.~\ref{Fig:2}, in which a gluon with momentum $p_{5}$ is emitted by the quark
$a$ that fragments into a photon with momentum $p_{\gamma}$. We denote by $p_{f}$ the
momentum of the collinear hadronic fragments of the quark $a$. The gluon may or may
not belong to the isolation cone around the photon, thus leading to two different
restrictions on the allowed kinematics of gluon 5.

\vspace{0.3cm}

\noindent
i) The gluon is outside the cone. We only have a condition on $p_{T \, f}$
\begin{equation}
\label{3.3}
p_{T \, f} \leq \varepsilon_{h} \ p_{ T \, \gamma} 
\end{equation} 

\noindent Defining 
\begin{equation}
\label{3.4}
p_{T \, a} =  p_{T \, f} + p_{T \, \gamma} \;, 
\;\;\;\;\;
z =  \frac{p_{ T \, \gamma}}{p_{T \, a}} \;,
\end{equation}

\noindent we write eq.~(\ref{3.3}) as 
\begin{equation}
\label{3.5}
z \geq z_{c} \;.
\end{equation}

\noindent 
The variable $z$ is the fragmentation variable appearing in the
fragmentation function $D_{q}^{\gamma}(z,M_{F})$. Going back to expressions
(\ref{2.5}) and (\ref{2.7}), we see that condition (\ref{3.3})
restricts the $z$-integration range. The lower limit now is $z_{c}$ instead of
$z_{min}$ (we are assuming that $z_{c}> z_{min}$, as is the case in the kinematical
configurations of interest in experiments at high-energy colliders). 

\vspace{0.3cm}

\noindent
ii) The gluon belongs to the cone. The isolation criterion (\ref{criterion})
implies
\begin{equation}
p_{T \, f} + p_{T \, 5} \leq \varepsilon_{h} \, p_{ T \, \gamma} \;,
\end{equation}
or, equivalently, 
\begin{equation}
\label{3.6}
p_{T \, a} + p_{T \, 5} \leq \frac{p_{ T \, \gamma }}{z_{c}} \;.
\end{equation} 

\noindent 
In terms of the variable $z$ in eq.~(\ref{3.4}) and of the variable $x$,
\begin{equation}
\label{3.7}
x = {p_{T \, a} \over p_{T \, a} + p_{T \, 5}} \;,
\end{equation} 

\noindent
eq.~(\ref{3.6}) is written as

\begin{equation}
\label{3.8}
z \ x \geq z_{c} \;.
\end{equation} 

\noindent
The constraint in eq.~(\ref{3.8}) means that, when the gluon is inside the cone,
we must also
restrict the $x$-integration range, so that 
$x \geq z_{c}/z$ instead of $x \geq z_{min}/z$ has to be used in eq.~(\ref{2.7}).

\vspace{0.3cm}

\noindent
In summary, we obtain the fragmentation component with isolation by
sub\-trac\-ting from the inclusive one the contribution that
violates eqs.~(\ref{3.5}) and (\ref{3.8}). Condition (\ref{3.5}) is
straightforwardly implemented by changing the $z$-integration range in
eqs.~(\ref{2.5}) and (\ref{2.7}). Condition (\ref{3.8}) modifies the $x$-integration
range and the function $K^{brems}_{HO}$
in eq.~(\ref{2.7}). Working again within the small-$R$ approximation, we can
perform a calculation similar to the one carried out in sect.~\ref{ancalc_incl}
in the case of the direct component. We thus
obtain the part of the higher-order correction that has to 
be subtracted from eq.~(\ref{2.7}) (after having implemented the constraint
$z>z_c$) when $R \ll 1$:
\begin{eqnarray}
\label{3.9} 
d\sigma_{\gamma jet}^{brems,sub} & = & 
      2 \pi \left( \frac{\alpha_{s}(\mu)}{\pi} \right) ^{3} 
      \sum_{b \atop{a,c = q,\bar{q}}} \int_{z_{c}}^{1} 
       D_{a}^{\gamma}(z,M_{F}) \, dz \nonumber \\
&   & \times \;\; \int_{\frac{z_{min}}{z}}^{\frac{z_{c}}{z}} \;dx \; 
     C_{F} 
     \left\{      
     \ln \left( \frac{R \ 
     p_{T \, \gamma} \ (1-x)}{z \ M_{F}} \right) 
     \left( \frac{1 + x^{2}}{1 - x} \right) + \frac{1}{2}(1-x) 
     \right\} \nonumber \\
&  & \times \;\; E_{c} 
\, \frac{d\widehat{\sigma}_{Born}({\scriptstyle A + B \to c + b + X})}
           {d \vec{p}_{c} \ d\eta_{b}} \ 
        d\eta_{b} \ d\eta_{c} \ p_{T \, c} \ dp_{T \, c} \;.
\end{eqnarray}  
\noindent
Here the photon momentum is related to the momentum of the parton $c$ by
$p_\gamma=zxp_c$.
As in the case of the direct component, the explicit $R$-dependence of the
contribution in the curly bracket can be eliminated
by choosing the scale $M_{F} = R \, p_{T \, \gamma}$, at the price
of introducing $R$-dependent effects through the $M_{F}$-dependence of the 
fragmentation function $D_{a}^{\gamma}(z,M_{F})$.

\vspace{0.3cm}

\noindent
In the above presentation, we have explicitly considered the subprocess in 
fig.~\ref{Fig:2},
which is
collinear- (when $R \to 0$) and infrared- (when $z_c \to 1$) divergent and 
thus leads to the most important HO
corrections. The same approach applies to other subprocesses. For instance, parton
$a$ can be a gluon and parton 5 a quark. We obtain corrections similar to those
in eq.~(\ref{3.9}), apart from the following replacement. The contribution inside the
curly bracket on the right-hand side of eq.~(\ref{3.9}) is:
\begin{equation}\label{crossings}
 \ln \left( \frac{R \ p_{T \, \gamma} \ (1-x)}{z \ M_{F}} \right) \,
 \widehat{P}_{qq}^{(4)}(x) + \frac{1}{2} \, P_{qq}^{(n-4)}(x) \;,
\end{equation}

\noindent
where the terms
\[
\widehat{P}^{(4)}_{qq}(x) =  C_{F} \, \frac{1+x^{2}}{1-x} \;, \;\;\;\;
         P_{qq}^{(n-4)}(x)=  C_{F} \, (1-x) \;,
\]
build the unregularized Altarelli--Parisi kernel in $n$ dimensions
$\widehat{P}^{(n)}_{qq}(x)=\widehat{P}^{(4)}_{qq}(x) - \varepsilon \, P_{qq}^{(n-4)}(x)$. 
The HO corrections from the subprocess with generic parton species $c$ and $a$
in fig.~\ref{Fig:2} are obtained from eq.~(\ref{3.9}) through the substitution
\begin{eqnarray*}
\widehat{P}^{(4)}_{qq}(x) & \rightarrow & \widehat{P}^{(4)}_{ac}(x) \;,\\
  P_{qq}^{(n-4)}(x) & \rightarrow &         P_{ac}^{(n-4)}(x) \;.
\end{eqnarray*}

\noindent
The explicit expressions of $\widehat{P}^{(4)}_{ac}(x)$ and
$P_{ac}^{(n-4)}(x)$ can be found, for instance, in ref.~\cite{bgpw}.

\vspace{0.3cm}

\noindent
The calculation described in this subsection is valid in the limit $R \ll 1$
and thus neglects corrections of ${\cal O}(R^{2n})$. These corrections are 
discussed in sect.~\ref{numcalc}.

\mysection{Numerical calculation and ${\cal O}(R^{2n})$ terms}\label{numcalc}

The isolation criterion (\ref{criterion}) modifies the calculation of
the HO corrections to prompt-photon production
by restricting the available phase space to final-state radiation.
In sect.~\ref{ancalc_isol}, we
have implemented these cuts by working in the collinear approximation for
pedagogical purposes. To go beyond the collinear approximation and keep
the complete $R$-dependence of the cross section, we have implemented all the
LO and NLO contributions in a computer programme, according to a combined
analytical and Monte Carlo approach. In this section we first briefly describe
the programme, then we present the results of our numerical
study.

\subsection{Brief presentation of the programme}
 
The code we use is derived from the NLO Monte Carlo programme DIPHOX 
\cite{bgpw}, designed to calculate the double-inclusive cross sections 
\[
E \frac{d \sigma ({\scriptstyle A + B \to F_{1} + F_{2} + X})}
       {d\vec{p}_{1} \ d\eta_{2}}
\]
and associated distributions, where $F_{1}$ and $F_{2}$ are large-$p_{T}$
particles, photons or hadrons \cite{bgpw,cfg}.
This programme
combines the phase-space slicing method \cite{slicing, slice} and the subtraction
method \cite{submeth, dipmeth} to treat the soft and collinear
singular parts of the perturbative matrix elements. 

\vspace{0.3cm}
 
\noindent
For a generic parton subprocess $1 + 2 \rightarrow 3 + 4 + 5$, the photon and one 
outgoing parton, say 3 and 4, have a high $p_{T}$ and are well separated in
phase space, while the remaining final-state parton, say 5, can be either soft
or collinear to one of the other four partons.
The phase space is sliced by using two arbitrary, unphysical
parameters $p_{Tm}$ and $R_{Th}$, with $p_{Tm} \ll |\vec{p}_{T\;3,4}|$ and 
$R_{Th} \ll 1$, in the following way:

\begin{itemize}
\item[-] Part I corresponds to $|\vec{p}_{T\;5}| < p_{Tm}$. This
cylinder supports the soft and initial-state collinear singularities. It
also yields a small fraction of the final-state collinear singularities
from the subregion in which parton 5 is very soft. 

\item[-] Part II.$a$ corresponds to the region where $|\vec{p}_{T\;5}|
\geq p_{Tm}$ and the parton~$5$
is inside a cone $C_3$ about the direction of 
particle~$3$, defined by $(y_5-y_3)^2+(\phi_5-\phi_3)^2~\leq~R_{th}^2$. 
This region
supports the final-state collinear singularities appearing when $5$ is 
collinear to $3$.

\item[-] Part II.$b$ is defined in a similar way as II.$a$,  but with 
the replacement of particle 3 by particle 4. The corresponding cone 
$C_4$ supports the final-state collinear singularities appearing when $5$ is 
collinear to $4$.

\item[-] Part II.$c$ is the remaining region:
$|\vec{p}_{T\;5}| \geq p_{Tm}$, and $\vec{p}_{T\;5}$ belongs to neither of 
the two cones $C_3$, $C_4$. This slice yields no divergences, and can thus be 
treated directly in four space-time dimensions.
\end{itemize}

\noindent
Collinear and soft singularities appear in parts I, II.$a$ and II.$b$. They are
first regularized by dimensional continuation from $4$ to $n = 4 - 2
\varepsilon$ ($\varepsilon < 0$) space-time dimensions. Then, the $n$-dimensional integration over 
the kinematic variables (transverse momentum, rapidity and azimuthal angles) of 
parton 5 is performed analytically over the phase-space regions in I, II.$a$ 
and II.$b$. 
After combination with the corresponding virtual contributions, the soft
singularities cancel, and the remaining collinear singularities that do not
cancel are factorized and absorbed in the parton distributions and fragmentation
functions. The resulting quantities correspond to pseudo cross sections, where
the hard partons in the  regions I, II.$a$ and II.$b$ are unresolved from the soft or
collinear parton 5, which has been `integrated out' inclusively on these
parts. The word `pseudo' means that they are not genuine cross sections, as
they are not positive-definite in general. These contributions as well as the one from
region II.$c$ are then encoded in the Monte Carlo computer programme.

\vspace{0.3cm}

\noindent
The integration over the phase-space region in II.$c$, 
which yields no divergences, is perfomed
numerically without any approximation. The implementation of the isolation 
criterion is straightforward: conditions (\ref{3.5}) or 
(\ref{3.8}) just cut the numerical phase-space integration 
when parton 5
is outside the isolation cone or inside the annulus
$R_{th}^{2} \leq 
\left( y_5 - y_{\gamma} \right)^{2} + \left(  \phi_5 - \phi_{\gamma}
\right)^{2}  \leq R^{2}$, respectively.

\vspace{0.3cm}

\noindent
The region in 
part I is treated according to the phase-space
slicing method. The integration is carried out analytically by neglecting terms
that are proportional to powers of $p_{Tm}$ and thus vanish when $p_{Tm} \to 0$.
This approximation implies that, when using the numerical programme, the
unphysical parameter $p_{Tm}$ has to be chosen sufficiently small
for the results to be independent of $p_{Tm}$.
In ref.~\cite{bgpw} 
it was checked that $p_{Tm}$ values of the order of half a per cent
of the minimum of $p_{T3}$ and $p_{T4}$ are sufficient.

\vspace{0.3cm}

\noindent
In principle, the dependence on the unphysical parameter $R_{th}$ can be
treated as that on $p_{Tm}$. However, such a procedure would lead to 
numerical instabilities when the radius $R$ of the isolation cone is also small.
Therefore, to avoid any approximation of the $R_{th}$-dependence,
the integration over the phase-space regions in part II.$a$ 
and II.$b$ is performed by using the subtraction method. The integrand (i.e. the
square of the matrix element) is first replaced by its collinear approximation
(see sect.~\ref{ancalc_incl}), and the corresponding integration is performed
analytically. This calculation is formally identical to the one
presented in sects.~\ref{ancalc_incl} and \ref{ancalc_isol}, the isolation-cone
radius $R$ now being replaced by the slicing parameter $R_{th}$. 
Therefore, after factorization of the collinear singularities in the
fragmentation functions,
the numerical programme contains expressions like
(cf. eq.~(\ref{3.9}))
\begin{eqnarray*}
&   & 2 \pi \left( \frac{\alpha_{s}(\mu)}{\pi} \right) ^{3}
     \sum_{b \atop{a,c = q,\bar{q}}}  
     \int_{z_{min}}^{1} D_{a}^{\gamma}(z,M_{F}) \ dz 
     \int_{\frac{z_{min}}{z}}^{1}      E_{c} 
      \frac{d\widehat{\sigma}_{Born}({\scriptstyle A + B \to b + c + X})}
           {d \vec{p}_{c} \ d\eta_{b}} \ 
      d\eta_{b} \ d\eta_{c} \ p_{T \, c} \ dp_{T \, c} \\  
&   & \times\;\; {C_{F}} 
\left\{ 
     \ln \left( \frac{R_{th} p_{T \, c}}{M_{F}} \right) 
         \left( \frac{1 + x^{2}}{1 - x} \right)_{+}  
     + (1 + x^{2}) \left( \frac{\ln (1-x)}{1-x} \right) _{+}
+ \frac{1 + x^{2}}{1-x} \ln x + \frac{1}{2}(1-x) 
\right\} dx \;,
\end{eqnarray*}  
in which the isolation criterion is simply implemented by changing
everywhere $z_{min}$ by $z_{c}$. Then, to take into account the exact
dependence on $R_{th}$, the difference between the exact matrix element
and its collinear approximation is integrated numerically, yielding
finite contributions (which are of ${\cal O}(R_{th}^{2n})$ when 
$R_{th} \to 0$) that are included in the Monte Carlo programme. The results,
therefore, do not depend on $R_{th}$.

In the following we present some numerical studies that implement the isolation
criterion in eq.~(\ref{criterion}). However, our NLO Monte Carlo programme can 
also be used to implement other definitions of isolation, such as, 
for instance, the isolation criterion in eq.~(\ref{d0criterion}).

\subsection{
Numerical results}

We start the presentation of our numerical investigations by studying the sensitivity of 
the cross section with respect to variations of the isolation parameters. We then examine the
dependence of the NLO results on the factorization and renormalization scale.
We also compare our complete numerical treatment with that in the 
collinear approximation. Finally, we study the isolated cross section as a
function of the transverse momentum of the photon.

\subsubsection{Sensitivity to the isolation parameters}

We use exact NLO expressions  for the isolated cross sections, where
`exact' means that the full $R$-dependence is kept, i.e. all the terms
proportional to $R^{2n}(n \geq 1)$ are taken into account on top of the collinear
approximation used in sect.~\ref{ancalc_isol}. 
We choose kinematical parameters
corresponding to those used by 
the CDF experiment at Tevatron Run Ib \cite{cdf}:
$\sqrt{S} = 1.8$~TeV, 
$-0.9 \leq y_\gamma \leq 0.9$ and $p_{T\,\gamma} = 15$~GeV, which corresponds to the lower range of
the photon  $p_{T}$ spectrum. We consider the prompt-photon inclusive cross
section

\[ 
\frac{d\sigma}{dp_{T\,\gamma}} = 
\int_{y_{min}}^{y_{max}} dy \ \frac{d\sigma}{dy_{\gamma} \ dp_{T\,\gamma}} \;\;.
\]
Similar studies can be done for photon--jet cross sections 
\cite{jet-photon}. We use the NLO
parton distribution functions of the set MRST-99
\cite{Martin:1999ww}, and the NLO fragmentation functions of set~II in
Bourhis et al. \cite{bfg}. The
calculations are done with $N_{f} = 5$ flavours.
The renormalization and
factorization scales $\mu$ and $M$ are both set equal to $p_{T\,\gamma}/2$. 

\vspace{0.3cm}

\noindent
Table~1 shows the sensitivity of the cross section to the value $R$ of the
isolation cone.
In this study we fixed $\varepsilon_{h} = 2/15 \simeq 0.13333$, which 
means that all events with hadronic transverse energy larger than 2~GeV in 
the isolation cone are rejected. The results without isolation are also reported
for comparison.
We verify that the Born cross sections are not sensitive to the isolation 
radius, as they should. 

\vspace{0.7cm}

\begin{center}
\begin{tabular}{|c|l|c|l|c|c|}
\hline
{Isolation radius} &\multicolumn{2}{|c|}{Direct contribution} &
\multicolumn{2}{|c|}{Fragmentation contribution} &{Total}\\
\hline
{R} &{Born} &{NLO} &{Born} &{NLO} &{NLO} \\
\hline
 1.0 &1764.6  &3318.4  &\ \ 265.0  &\ \ 446.7  &3765.1  \\
 0.7 &1764.6  &3603.0  &\ \ 265.0  &\ \ 495.0  &4098.0  \\
 0.4 &1764.6  &3968.9  &\ \ 265.0  &\ \ 555.6  &4524.5  \\
 0.1 &1764.6  &4758.2  &\ \ 265.0  &\ \ 678.9  &5431.1  \\
& & & & & \\
Without isolation &1764.6  &3341.1  &1724.3  &1876.8  &5217.9  \\
\hline
\end{tabular}

\vspace{0.3cm}

Table 1. Isolated cross sections (the values are given in pb/GeV)
corresponding to $\varepsilon_h = 0.13333$.
\end{center}

\vspace{0.3cm}

\noindent
It is interesting to note that the HO contributions, both to the direct and to
the fragmentation components, increase when $R$ decreases. This is due to the fact that the
implementation of isolation amounts to subtracting a contribution proportional to
$\ln R$ from the non-isolated cross section (see eqs.~(\ref{3.2}) and 
(\ref{3.9})). Since this subtracted contribution
is negative when $R < 1$, the HO contribution to the direct component of the
isolated cross section
is quite large for small values of $R$. A similar behaviour is
observed in the HO contribution to the fragmentation component. When all contributions are
taken into account, the total cross section (direct +
fragmentation) strongly increases with decreasing $R$. 

\vspace{0.3cm}

\noindent
In particular, when $R=0.1$, the NLO calculation gives 
an unphysical result: the isolated cross section turns out to be larger than 
the non-isolated one! Such a behaviour had to be expected in view of the
discussion at the end of sect.~\ref{dirwis}. The NLO results in 
table~1 imply that the value $R \sim 0.1$ is sufficiently small to
demand the inclusion of beyond-NLO perturbative terms and non-perturbative
contributions.

\vspace{0.3cm}

\noindent
The sensitivity of the cross sections to variations of $\varepsilon_{h}$ 
is displayed in table~2. Now we fix $R = 0.7$.  

\vspace{0.7cm}

\begin{center}
\begin{tabular}{|c|l|c|l|c|c|}
\hline
{Energy cut} &\multicolumn{2}{|c|}{Direct contribution} &
\multicolumn{2}{|c|}{Fragmentation contribution} &{Total}\\
\hline
$\varepsilon_h$&{Born}&{NLO}&   {Born}    & {NLO} &{NLO}
\\ \hline
0.03333  &  1764.6  &3820.9 &\ \ \, 60.3 &\ \ 168.7  &3989.6  \\
0.06667  &  1764.6  &3734.3 &\ \  135.2 &\ \ 303.4  &4037.7  \\
0.13333  &  1764.6  &3603.0 &\ \  265.0 &\ \ 495.0  &4098.0  \\
0.33333  &  1764.6  &3434.1 &\ \  571.2 &\ \ 883.9  &4318.0  \\
0.66667  &  1764.6  &3359.5 &\ \  930.4 &\  1307.5  &4667.0  \\
1.00000  &  1764.6  &3340.4 &\   1173.1 &\  1579.4  &4919.8  \\
\hline
\end{tabular}

\vspace{0.3cm}

Table 2. Isolated cross sections (the values are given in pb/GeV)
corresponding to $R =0.7$.
\end{center}

\vspace{0.3cm}

\noindent
Note that, already at the Born level,  
the fragmentation component 
is quite sensitive to $\varepsilon_{h}$ and strongly decreases when 
such a cut is installed.

\vspace{0.3cm}

\noindent
The ratios NLO/Born 
increase when $\varepsilon_{h}$
decreases, indicating that the effect of higher-order corrections is larger 
at small than at large $\varepsilon_{h}$. This is due to
the following mechanism. Radiation collinear to the photon is 
more suppressed by the transverse-energy isolation cut than hard non-collinear
radiation. Since the collinear contributions are negative when evaluated in the
${\overline{MS}}$ factorization scheme (see e.g. eq.~(\ref{2.18})),
their strong
suppression leads to a sizeable NLO correction when $\varepsilon_{h}$ decreases.
We also note that, as expected,
the effect of the $\varepsilon_{h}$ cut-off is very large on the fragmentation 
component, in which a large part of the $z$-integration domain is suppressed.
The total (direct + fragmentation) cross section at NLO is rather stable
with respect to $\varepsilon_{h}$ variations, because of the behaviour of the
direct contribution.

\vspace{0.3cm}

\noindent
We point out that there is no infrared divergence coming from the NLO collinear
contribution when $\varepsilon_{h} \to 0$ $(z_c \to 1)$. The direct
component contains a term proportional to 
$\int_{z_{c}}^{1} dz \ \ln(1 - z) \simeq (1 - z_{c}) \ln(1 - z_{c})$ when 
$z_{c} \to 1$ (see eqs.~(\ref{2.18}) and (\ref{3.2})). 
The fragmentation component involves terms 
proportional to $(1 - z_{c}) \ln^{3} (1 - z_{c})$: a factor $\ln^2(1 - z_c)$
comes from integrating the behaviour $\ln(1 - x)/(1- x)$ in
eq.~(\ref{3.9}), and a factor $(1 - z_{c}) \ln(1 - z_{c})$ comes from the
convolution with the large-$z$ behaviour, proportional to $\ln(1 - z)$, 
of the NLO fragmentation function $D_{q}^{\gamma}(z,M_{F})$ \cite{bfg}.
Still in both cases the NLO cross section is finite when
$\varepsilon_{h} \to 0$. However, there are infrared-divergent
contributions that are not smoothed out by the convolution with the
fragmentation function $D_{q}^{\gamma}(z,M_{F})$. They correspond to 
soft gluons emitted non-collinearly to the photon (for instance, emitted
from the initial-state quark of the LO direct 
subprocess $q+{\bar q} \to \gamma +g$)
and produce terms proportional to $R^{2} \ln \varepsilon_{h}$ in the direct
component of the NLO cross section \cite{gordon-vogelsang2}. The effect of these
terms is thus suppressed when the size of the isolation cone is relatively
small. The numerical stability of the NLO results in table~3 suggests that
the contribution of infrared-divergent terms is not dominant, unless
the parameter $\varepsilon_{h}$ becomes very small.

\subsubsection{Scale dependence}

Until now all calculations have been performed with the renormalization scale
$\mu$, the initial-state factorization scale $M$ and final-state fragmentation 
scale $M_{F}$, all equal to $p_{T\, \gamma}/2$. Here we study the sensitivity of the
isolated cross section with respect to scale variations. Choosing standard
isolation parameters, $\varepsilon_{h} = 2/15$ and $R = 0.7$, we vary the scale
$\mu = M = M_{F}$ between $p_{T\, \gamma}/2$ and $2 p_{T\, \gamma}$. The results are given in
table~3. They can be compared with the corresponding results obtained without 
isolation, which are displayed in table~4.

\begin{center}
\begin{tabular}{|c|c|c|c|}
\hline
{Scale}&{Direct contribution}&{Fragmentation contribution}&{Total}\\ \hline
         & {(NLO)} & {(NLO)} &       {(NLO)}               \\ \hline
$p_{T}/2$&   3603.0     &    495.0     &  4098.0            \\
$p_{T}$  &   3155.3     &    576.2     &  3731.5            \\
$2p_{T}$ &   2840.7     &    631.7     &  3472.4            \\ \hline
\end{tabular}
\vspace{0.3cm}

Table 3. Scale dependence of the isolated cross section in pb/GeV.
\end{center}

\begin{center}
\begin{tabular}{|c|c|c|c|}
\hline
{Scale}&{Direct contribution}&{Fragmentation contribution}&{Total}\\ \hline
         &  {(NLO)}   &       {(NLO)} &    {(NLO)}         \\ \hline
$p_{T}/2$&    3341.1       &      1876.8        &  5217.9   \\
$p_{T}$  &    2643.7       &      2122.6        &  4766.3   \\
$2p_{T}$ &    2188.8       &      2250.7        &  4439.5   \\ \hline
\end{tabular}

\vspace{0.3cm}

Table 4. Scale dependence of the non-isolated cross section in pb/GeV.
\end{center}

\noindent
The cross section decreases monotonically in both cases when
the scale increases. The magnitude of the effect is comparable in the two cases:
a relative variation of the cross section by $\sim 16\%$ is observed in the range 
of scales considered. We conclude that the implementation of isolation does not 
induce any extra sensitivity to the scale arbitrariness, at least when the three
scales $\mu,M,M_{F}$ are kept equal. 

\vspace{0.3cm}

\subsubsection{Comparison with the collinear approximation} 

The isolation criterion used in 
ref.~\cite{gordon-vogelsang2} is based on an upper limit on the hadronic 
c.m. energy rather than on the hadronic transverse energy
inside the isolation cone. As pointed out in ref.~\cite{gordon-vogelsang2} and
discussed in general terms in sect.~\ref{factorisation},
in the case of soft and collinear parton radiation, the two variants of
the isolation criterion become equivalent by simply identifying
the c.m.-energy parameter $\varepsilon$ of 
ref.~\cite{gordon-vogelsang2} with our transverse-energy parameter
$\varepsilon_{h}$ in eq.~(\ref{epsilon-h}). Since the approximate
NLO calculation performed in ref.~\cite{gordon-vogelsang2} is actually
based on the soft and collinear limits, it can thus be compared with our
calculation by setting $\varepsilon_{h}=\varepsilon$.

\vspace{0.3cm}

\noindent
More precisely, 
the authors of ref.~\cite{gordon-vogelsang2} derived approximate analytical
expressions for the NLO terms $\sigma_{HO}^{{\rm is}}$ in eqs.~(\ref{isa}) and
(\ref{isb}). These expressions are valid in the small-cone\footnote{In 
ref.~\cite{gordon-vogelsang2} the cone size $\delta$ is defined with respect to
the relative angles in the c.m. frame. When $R \ll 1$, we thus have
$R=\delta \cosh \eta_\gamma$.} approximation 
$R \ll 1$, and thus neglect corrections that are of ${\cal O}(R^{2})$ when
$R \rightarrow 0$. In this limit the collinear approximation is valid, and
our results in eqs.~(\ref{3.2}) and (\ref{3.9}) fully agree with those in
ref.~\cite{gordon-vogelsang2}. In the case of the direct contribution
$\sigma_{HO}^{\gamma,{\rm is}}$, the authors of ref.~\cite{gordon-vogelsang2}
also computed the dominant correction due to soft-gluon emission.  
This correction behaves as $R^{2} \ln \varepsilon_{h}$ when
$\varepsilon_{h} \to 0$ and $R \ll 1$.
 
\vspace{0.5cm}

\begin{center}
\begin{tabular}{|c|c|c|c|c|}
\hline
{R}&\multicolumn{2}{|c|}{$\varepsilon_{h}= 2/30 \simeq 0.06667$}
   &\multicolumn{2}{|c|}{$\varepsilon_{h}= 2/15 \simeq 0.13333$}\\ \hline
   &{Small-cone approx.}&{No approx.}&{Small-cone approx.}
   &{No approx.}
\\ \hline
 1.0 &     292.5   &   267.7  &     482.4    &  446.7  \\
\ .7 &     313.7   &   303.4  &     512.8    &  495.0  \\
\ .4 &     344.0   &	 345.6  &     560.9    &  555.6   \\
\ .1 &     431.8   &	 432.9  &     678.9    &  678.9  \\
\hline
\end{tabular}

\vspace{0.3cm}

Table 5. Comparison between our results (no approximation) and the small-cone
            approximation \cite{gordon-vogelsang2}
	    for the fragmentation component. The cross sections are given in
	    pb/GeV.
\end{center}

\vspace{0.3cm}

\noindent
In the case of the fragmentation component, Table~5 shows a comparison
between our NLO calculation and that of ref.~\cite{gordon-vogelsang2}.
We see that there is a good agreement between the two calculations when $R \ll 1$,
the collinear approximation becoming more accurate as $R$ decreases (our results
are obtained by Monte Carlo integration with an accuracy of $1\%$). For large
values of $R \sim 1$, the small-cone approximation overestimates the
fragmentation contribution to the isolated cross section by about 8\%. 
In the case of the direct component, similar results were found in 
ref.~\cite{gordon-vogelsang2} by a numerical comparison with the 
calculation performed by a Monte Carlo code. 

\begin{figure}[p]
\includegraphics[scale=0.8]{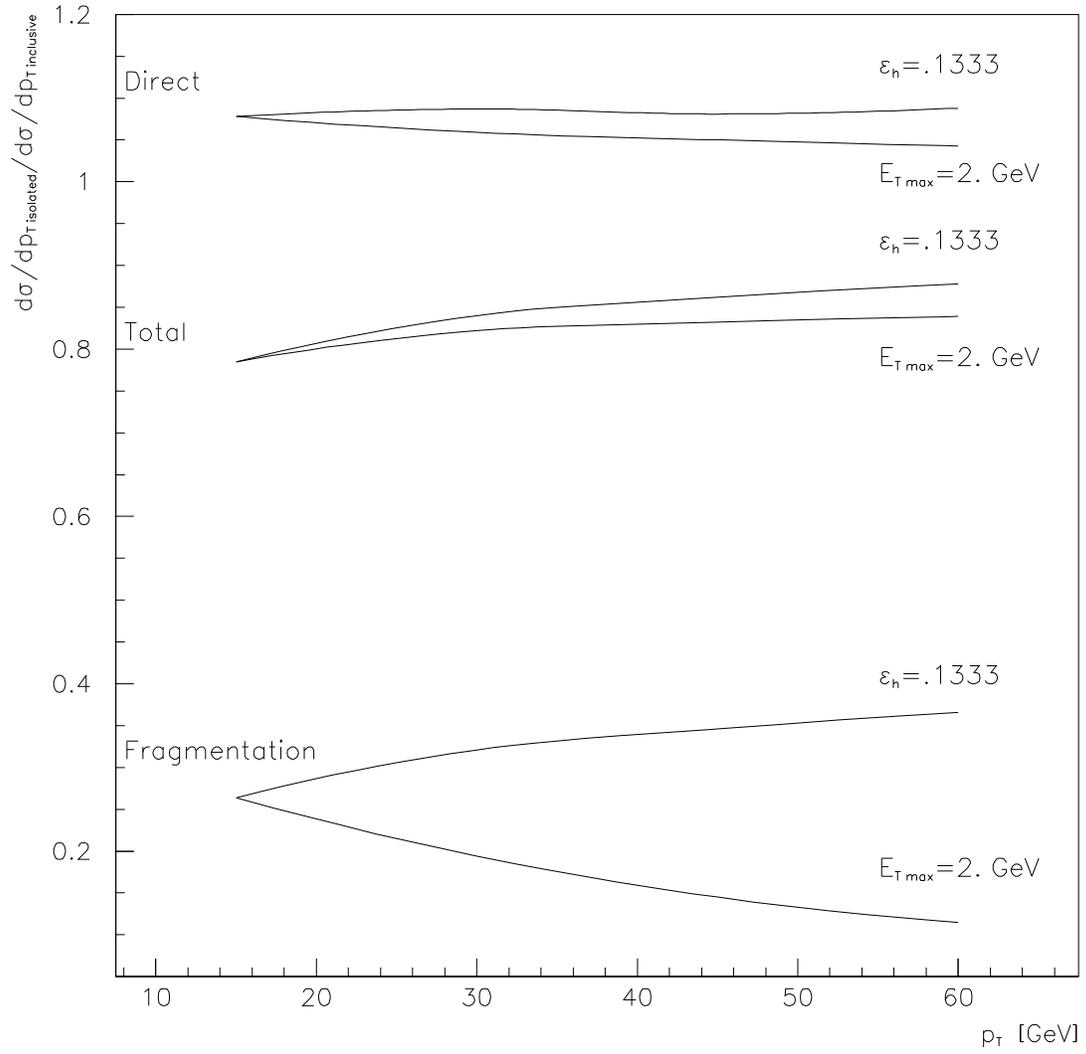}
\caption{\label{Fig:ptdepen}{\em Ratios `isolated'/`non-isolated' for the
direct contribution, the fragmentation contribution and the total contribution
to the cross section $d \sigma/dp_{T \, \gamma}$ at NLO.}}
\end{figure}

\subsubsection{Effect of the isolation as a function of $p_{T \, \gamma}$}

Finally we study the effect of isolation as a function of 
$p_{T \, \gamma}$. Figure~\ref{Fig:ptdepen} shows the ratios 
`isolated'/`non-isolated' of the direct contributions, the
fragmentation contributions, and the total contributions to the NLO cross sections 
$d \sigma/d p_{T \, \gamma}$.
The ratios 
are displayed in the range 15~GeV $ \leq p_{T \, \gamma} \leq$ 60~GeV, 
for two different choices of isolation parameters, both with an isolation cone of radius $R = 0.7$:
\begin{itemize}
\item[] one with a fixed value of $E_{T \, max}= 2$~GeV,
\item[] one with a fixed value of 
$\varepsilon_{h}= E_{T \, max}/p_{T \, \gamma}=2/15\simeq 0.1333$,
\end{itemize}
which coincide when $p_{T \, \gamma} = 15$~GeV.

\vspace{0.2cm}

\noindent
As for the fragmentation component,
isolation using a fixed $E_{T \, max}$ acts more and more severely as 
$p_{T \, \gamma}$ increases, while by fixing
$\varepsilon_{h}$ the amount of accompanying
transverse energy in the cone increases with  $p_{T \, \gamma}$. Therefore 
the ratios `isolated'/`non-isolated' with fixed $E_{T \, max}$ and with
fixed $\varepsilon_{h}$ have opposite variations when $p_{T \, \gamma}$
varies.

\vspace{0.2cm}

\noindent
Isolation has a rather small effect on the direct contribution in both cases,
since it does not act at the Born level.

\vspace{0.2cm}

\noindent
The effect of isolation on the total contribution to the NLO
cross section slightly decreases as 
$p_{T \, \gamma}$ increases, but it actually depends only weakly on 
$p_{T \, \gamma}$.

\section{Conclusions and
outlook}
\label{conclusion}

\subsection{Conclusions}\label{recap}

In this paper we have considered the production of isolated photons in hadronic
collisions. We have shown that isolation does not spoil the factorization
properties that are valid in the case of inclusive production, and we have
discussed the general factorized structure of the isolated-photon cross section.
We have then performed a detailed NLO study of the implementation of the
isolation criterion in the direct and fragmentation components of
the production cross section. For pedagogical purposes, we have first 
given analytic expressions of the isolated cross sections obtained by using the
collinear approximation, i.e. valid in the limit of small radius $R$ of the
isolation cone. To go beyond the collinear approximation, we have 
implemented the isolation criterion in a computer programme that 
encodes the exact dependence of the cross section on the
cone radius at NLO accuracy.

\vspace{0.3cm}

\noindent
The main result of our NLO numerical study is that the inclusive cross section for 
isolated photons has a magnitude comparable to the one for non-isolated 
photons. Each separate component, direct and fragmentation, 
sizeably
depends on the isolation parameters, both at LO and NLO.
However, the measurable cross section, given by the sum  `direct + fragmentation',  does
not vary by more than 10\% when $\varepsilon_{h}$ varies between $2/30 
\simeq 0.06667$ and 0.333, or when $R$ varies between 0.4 and 0.7. Nonetheless,
with small cones such as $R \sim 0.1$, the result of the
NLO calculation for the isolated cross section becomes larger
than the one in the non-isolated case. This counterintuitive and unphysical 
result reflects the fact that the fixed-order perturbative calculation is no
longer reliable when $R$ is very small: when $\alpha_s \ln 1/R^2 \sim 1$,
perturbative contributions beyond the NLO have to be taken into account; when
$R \, p_{T \, \gamma} \sim {\cal O}(1$~GeV), also non-perturbative contributions
are demanded.
The scale dependence of the cross sections is similar
in the isolated and non-isolated
cases:
the cross sections decrease monotonically
by $\sim 16\%$ when the common scale $\mu=M=M_{F}$
varies between $p_{T}/2$ and $2 p_{T}$. Comparing our NLO calculation with
its small-cone approximation \cite{gordon-vogelsang2}, we find that the latter 
is adequate at small $R$ and overestimates
the cross section by about 8 to 10\% 
at large $R$ values, $R \sim 1$. Finally, isolation is
found to weakly affect the $p_{T \gamma}$-dependence of the cross section  
$d \sigma/d p_{T \, \gamma}$ in  the range 15~GeV $\leq p_{T \,
\gamma}\leq$ 60~GeV.

\subsection{Open problems}\label{nasty}

We conclude with some remarks on several questions left open by our NLO
treatment when the cone radius $R$ becomes too small,
as well as when the parameter $E_{T \, max}$ is small and isolation
becomes very tight.

\subsubsection{Small cone radius}

As  seen numerically in sect.~\ref{numcalc}, the result of the NLO calculation
for the isolated cross section at $p_{T \, \gamma} = 15$~GeV violates 
the physical constraint $\sigma({\rm with \ isol.}) < \sigma({\rm no \ isol.})$
when $R \sim 0.1$. This value of $R$ is smaller than the ones relevant to 
experimental practice.
However, the value $R \sim 0.1$ is not much 
smaller than the 
value $R = 0.4$ that is used in the most recent measurements at the Tevatron 
\cite{d0,cdfnew}
and that is going to be used at the  LHC \cite{atlascms}.
Therefore, we should be concerned about the actual value of $R$ below which the reliability of
the fixed-order perturbative calculation breaks down. 

\vspace{0.3cm}

\noindent
Note that $p_{T \, \gamma} = 15$~GeV and $R=0.1$ imply
$R \, p_{T \, \gamma} \sim 1$~GeV and $\alpha_s(\mu^2) \ln 1/R^2 \sim 1$
(recall that $\mu=p_{T \, \gamma}/2$ is used in the numerical results
of table~1). Therefore, as discussed at the end of sect.~\ref{dirwis}, the scale
$R \, p_{T \, \gamma}$ is close to the non-perturbative region and, at the same
time, higher-order
corrections proportional to $(\alpha_s \ln 1/R^2)^n$ can be relevant. 
Summation of the logarithmic
dependence on $R$ to all perturbative orders,
combined with a careful study of the border-line between
perturbative and non-perturbative regions, has to be undertaken  
to improve our understanding of the small-$R$ behaviour of isolated-photon
prodution. 
Work in this direction is in progress, and the results will be reported
elsewhere.

\subsubsection{Tight isolation}

Formally, the perturbative calculation of the cross section of isolated prompt
photons is infrared-divergent in the limit $\varepsilon_{h} \to 0$.
Indeed, $\varepsilon_{h} = 0$ would imply that the isolation cone 
about the photon would become an absolutely forbidden region of the phase space 
for gluon radiation, no matter how soft it is, thus spoiling
the cancellation of infrared 
singularities between real and virtual soft-gluon contributions.
As discussed in sect.~\ref{numcalc}, at NLO this divergent behaviour shows up
as a logarithmic term proportional to $\alpha_{s} R^{2} \ln \varepsilon_{h}$
in the direct component of the cross section.

\vspace{0.3cm}

\noindent
In practice, using a cone of radius $R =0.7$, we have found no significant 
infrared sensitivity in our 
numerical study, down to the very low 
value $\varepsilon_{h}= 0.033333$, which corresponds to 
$E_{T \, max} = 0.5$~GeV for a photon with $p_{T \, \gamma} = 15$~GeV.
Therefore the implementation of the isolation criterion 
(\ref{criterion}) with (finite but) tight transverse-energy cuts 
does not seem to destabilize the numerical convergence of
the perturbative expansion. 
Nonetheless, owing to the presence of higher powers of $\ln \varepsilon_{h}$
at higher perturbative orders, the actual sensitivity of the cross section
to very low values of $\varepsilon_{h}$ is probably 
underestimated in the present NLO calculation. 

\vspace{0.3cm}

\noindent
In the present work we have treated all the quarks as being massless.
This treatment is not adequate 
in the case of a heavy quark when $E_{T \, max}$ is 
comparable to its mass. 
This issue especially concerns heavy quarks that are experimentally not
identified.
In this case, better theoretical calculations and  
studies of heavy-quark fragmentation based on Monte Carlo event generators
are required.

\vspace{0.3cm}

\noindent
In the NLO calculation, isolation is implemented at the parton level.
In an actual event, a
fraction of the hadronic transverse energy that accompanies a photon to be
selected by the criterion (\ref{criterion}) comes also from the low-$p_{T}$ 
underlying event, as well as from pile-up effects in collisions at very high 
luminosity.
Part of these effects can be mimicked in the NLO calculation by using
an effective value of $E_{T \, max}$ that is lower than that imposed at the
detector level. However, this procedure does not take into account the fact
that the underlying event causes also some suppression of the direct component
at the Born level, which is independent of $E_{T \, max}$. Model estimates
\cite{fgh} of this suppression indicate that it can be quantitatively relevant 
when the value of $E_{T \, max}$ at the detector level is small.
In general, as the isolation becomes tight, the perturbative calculation has to be
supplemented by careful studies of the effects of the underlying event and
pile-up.

\vspace{0.5cm}

\noindent 
{\bf Aknowledgements.} We thank Werner Vogelsang for sending us his numerical 
results. 
We also thank Stefano Frixione and Bryan Webber for discussions.
This work was supported in part by the EU Fourth Training Programme
``Training and Mobility of Researchers", Network ``Quantum Chromodynamics and
the Deep Structure of Elementary Particles", Contract FMRX-CT98-0194 (DG 12 -
MIHT). LPT is a  ``Unit\'e Mixte de  Recherche du CNRS (UMR 8627) associ\'ee
\`a l'Universit\'e de Paris XI". LAPTH is a ``Unit\'e Mixte de  Recherche du 
CNRS (UMR 5108) associ\'ee \`a l'Universit\'e de Savoie".

\end{fmffile}

\end{document}